\documentclass[aps,prd,onecolumn,showpacs,showkeys,superscriptaddress,preprintnumbers,floatfix,reprint]{revtex4}
\usepackage{graphicx}
\usepackage{amsmath}
\usepackage{bm}
\usepackage{yhmath}
\usepackage{hyperref}
\usepackage{subfigure}
\usepackage{color}
\usepackage{cases}
\usepackage{subfigure}
\usepackage{dcolumn,booktabs,bm}

\usepackage{amsfonts,amssymb,stmaryrd,latexsym,amsmath}
\usepackage{textcomp}

\usepackage{array}
\newcounter{RomanNumber}

\newcommand{\lyxmathsym}[1]{\ifmmode\begingroup\def\b@ld{bold}
  \text{\ifx\math@version\b@ld\bfseries\fi#1}\endgroup\else#1\fi}

\def\arccosh{{\rm arccosh}}
\def\rmI{{\rm I}}
\def\rmII{{\rm I\!I}}
\def\rmIII{{\rm I\!I\!I}}

\def\mud{{\lambda}}

\allowdisplaybreaks

\begin{document}

\title{Decuplet to octet baryon transitions in chiral perturbation theory}

\author{Hao-Song Li}\email{haosongli@pku.edu.cn}\affiliation{School of Physics and State Key Laboratory of Nuclear Physics and Technology, Peking University, Beijing 100871, China}

\author{Zhan-Wei Liu}\email{liuzhanwei@lzu.edu.cn}
\affiliation{School of Physical Science and Technology, Lanzhou
University, Lanzhou 730000, China}

\author{Xiao-Lin Chen}\email{chenxl@pku.edu.cn}
\affiliation{School of Physics, Peking University, Beijing 100871,
China}

\author{Wei-Zhen Deng}\email{dwz@pku.edu.cn}
\affiliation{School of Physics, Peking University, Beijing 100871,
China}

\author{Shi-Lin Zhu}\email{zhusl@pku.edu.cn}\affiliation{School of Physics and State Key Laboratory of Nuclear Physics and Technology, Peking University, Beijing 100871, China}\affiliation{Collaborative Innovation Center of Quantum Matter, Beijing 100871, China}

\begin{abstract}

We have systematically investigated the decuplet (T) to
octet (B) baryon ($T\rightarrow B\gamma$) transition magnetic moments to the
next-to-next-to-leading order and electric quadruple moments to the
next-to-leading order in the framework of the heavy baryon chiral
perturbation theory. Our calculation includes the contributions from
both the intermediate decuplet and octet baryon states in the loops.
Our results show reasonably good convergence of the chiral expansion and
agreement with the experimental data.
The analytical expressions may be useful to the chiral extrapolation of the lattice
simulations of the decuplet electromagnetic properties.

\end{abstract}

\maketitle

\thispagestyle{empty}

\section{Introduction}\label{Sec1}

The electromagnetic property of baryons has been an important topic
in both theory and experiment. There have been extensive
investigations of the baryon magnetic moments \cite{Adkins:1983ya,Cohen:1986va,Leinweber:1990dv,Leinweber:1992hy,Jenkins:1992pi,Butler:1993ej,Banerjee:1995wz,
Meissner:1997hn,Zhu:1998aj,Meissner:1999hk,Zhu:2000gn,puglia2,Kubis:2000aa,Puglia:2000jy,Savage:2001dy,Arndt:2003we,
Cloet:2003jm,Gockeler:2003ay,Pascalutsa:2004je,Alexandrou:2006ru,Hacker:2006gu,Arrington:2006zm,Pascalutsa:2007wb,
Lin:2008mr,Alexandrou:2008bn,Alexandrou:2009hs,Geng:2009ys,
Cloet:2014rja,Shanahan:2014uka,Carrillo-Serrano:2016igi}.
On the other hand, the baryon decuplet to octet electromagnetic transition also
probes the inner structure and possible deformation of both the
decuplet and octet batyons. In the past several decades, there have
been many investigations of the transition properties from both
experimental and theoretical perspectives.

The model-independent analysis of the $T\rightarrow B\gamma$
transition amplitude was first performed in
Refs.~\cite{Bjorken:1966ij,Jones:1972ky}. From the spin-parity
selection rule, the $T\rightarrow B\gamma$ transition amplitudes
contain the magnetic dipole ($M1$), electric quadrupole ($E2$), and
Coulumb quadrupole ($C2$) contributions. In the quark-model picture,
a spin flip of a quark in the s-wave state leads to the $M1$ type of
transition, while any d-wave admixture in the octet or the decuplet
wave functions allows for the $E2$ and $C2$ transitions. Thus, the
E2 to M1 ratio $R_{\rm EM}$ is a signature of the d-wave components
and the deviation of the nucleon from spherical symmetry. One of the
first successes of the constituent quark model was the prediction of
the $\Delta(1232)\rightarrow N\gamma$ transition magnetic moment~\cite{Beg:1964nm}. If d-waves are included, the electromagnetic
ratios are nonzero~\cite{Becchi:1965zz}. Since then, more
sophisticated quark models have been developed to study the
$\Delta(1232)\rightarrow N\gamma$
transition~\cite{Isgur:1981yz,Gershtein:1981zf,Bourdeau:1987ih,Gogilidze87,
Hemmert:1994ky,Buchmann:2004ia,Ramalho:2008aa}. In
Ref.~\cite{Faessler:2006ky}, the authors studied the magnetic
moments of $\Delta(1232)\rightarrow N\gamma$
 transition with relativistic quark models improved by chiral corrections.

Although quark model is rather successful in predicting the internal
structure of the baryons, the $\Delta(1232)\rightarrow N\gamma$
transition magnetic moment is generally underestimated by 25\% in
the constituent quark model. The decuplet and octet baryons are
almost degenerate. Moreover, the decuplet baryon nearly entirely
decays into the Goldstone boson and octet. It is essential to
consider the Goldstone boson cloud effect.

Chiral perturbation theory (ChPT)~\cite{Weinberg:1978kz} is a very
useful framework to take into account the chiral corrections in
hadron physics in the low-energy regime. In
Ref.~\cite{Butler:1993ht}, the authors studied $T\rightarrow
B\gamma$ transition with heavy baryon chiral perturbation theory
(HBChPT)~\cite{Jenkins:1991} to next-to-leading order. In
Ref.~\cite{Gellas:1998wx}, the "small scale expansion"
(SSE)~\cite{Hemmert:19978} was used to calculate the transition form
factors to $\mathcal{O}(\epsilon^3)$ with two light mass scales
included: the pion mass and the octet and decuplet baryon mass
splitting. The decuplet to octet baryon electromagnetic transition
form factors have also been calculated in quenched and partially
quenched chiral perturbation theory in Ref.~\cite{Arndt:2003vd}. An
analysis of the electromagnetic transition current was presented to
$\mathcal{O}(\epsilon^3)$ in the framework of the non-relativistic
SSE chiral effective field theory in Ref.~\cite{Gail:2005gz}. In
Ref.~\cite{Pascalutsa:2005ts}, the authors performed a relativistic
chiral effective-field theory calculation of the pion
electroproduction off the nucleon reaction in the $\Delta(1232)$
resonance region.

Besides the quark models and ChPT, the $T\rightarrow B\gamma$
transition was also studied with various approaches such as the the
cloudy bag model~\cite{Kaelbermann:1983zb,Bermuth:1988ms,Lu:1996rj},
Skyrme model~\cite{Wirzba:1986sc,Abada:1995db,Walliser:1996ps}, QCD
sum rules~\cite{Wang:2009bh}, large $N_C$
limit~\cite{Jenkins:2002rj,Buchmann:2002mm} and lattice
QCD~\cite{Leinweber:1992pv,Alexandrou:2003ea,Alexandrou:2004xn,Ramalho:2009df}.
Especially, the electromagnetic transition moments of the baryon
octet to decuplet were first calculated with quenched lattice QCD
simulation in Ref.~\cite{Leinweber:1992pv}. The electromagnetic form
factors of the $\Delta(1232)\rightarrow N\gamma$ transition were
evaluated both in quenched lattice QCD and using two dynamical
Wilson fermions in Refs.~\cite{Alexandrou:2003ea,Alexandrou:2004xn}.
In Ref.~\cite{Ramalho:2009df}, the authors studied the valence quark
contributions to the $\Delta(1232)\rightarrow N\gamma$ transition in
the lattice QCD regime in the framework of the covariant spectator
formalism.

In this work, we will calculate the $T\rightarrow B\gamma$
transition amplitudes up to $\mathcal{O}(p^4)$ (or
$\mathcal{O}(\epsilon^4)$) and extract the transition magnetic
moments to $\mathcal{O}(p^3)$ in the framework of HBChPT. We
explicitly consider both the octet and decuplet intermediate states
in the loop calculation as the octet and decuplet baryons couple
strongly. We use the dimensional regularization and modified minimal
subtraction scheme to deal with the divergences from the loop
corrections. At last, we discuss the convergence of the chiral
expansion of the transition magnetic moments. We also systematically
calculate the electro quadrupole moments to next-to-leading order
and obtain the E2 to M1 ratio $R_{\rm EM}$ for the decuplet to octet
baryon transitions. We collect the M1 and E2 amplitudes and decay
width of the decuplet to octet baryon transitions in the
Appendix~\ref{appendix-C}.

This paper is organized as follows. In Section \ref{Sec3}, we
discuss the decuplet to octet baryon electromagnetic transition form
factors. We introduce the effective chiral Lagrangians of the
decuplet baryon in Section \ref{Sec2}. In Section
\ref{secFormalism}, we calculate the decuplet to octet baryon transition
magnetic moments order by order. We estimate the low-energy
constants and present our numerical results in Section \ref{Sec6}
and conclude in Section \ref{Sec7}. We collect some useful formulae
and the coefficients of the loop corrections in the
Appendix~\ref{appendix-A} and~\ref{appendix-B}.

\section{Decuplet to octet baryon electromagnetic transition form factors}\label{Sec3}

When the electromagnetic current is sandwiched between decuplet and
octet baryon states, one can write down the general matrix elements
which satisfy the gauge invariance, parity conservation and time
reversal invariance ~\cite{Jones:1972ky}:
\begin{equation}
<B(p)|J_{\mu}|T(p^{\prime})>=e\bar{u}(p)O_{\rho\mu}(p^{\prime},p)u^{\rho}(p^{\prime}),
\end{equation}
with
\begin{equation}
O_{\rho\mu}(p^{\prime},p)=\frac{G_{1}}{2M_B}(q_{\rho}\gamma_{\mu}-q\cdot\gamma
g_{\rho\mu})\gamma_{5}+\frac{G_{2}}{4M_{B}^{2}}\frac{1}{M_{B}+M_{T}}(q\cdot
Pg_{\rho\mu}-q_{\rho}P_{\mu})q\hspace{-0.5em}/\gamma_{5}.
\label{eq_new_current}
\end{equation}
where $p$ and $p'$ are the momenta of the octet and decuplet baryons. In
the above equations, $P=\frac{1}{2}(p^{\prime}+p)$,
$q=p^{\prime}-p$, $M_{B}$ is octet-baryon mass, $M_{T}$ is
decuplet-baryon mass, and $u_{\rho}(p)$ is the Rarita-Schwinger
spinor for an on-shell heavy baryon satisfying
$p^{\rho}u_{\rho}(p)=0$ and $\gamma^{\rho}u_{\rho}(p)=0$.

In the heavy baryon limit, the baryon field $B$ can be decomposed
into the large component $\mathcal{N}$ and the small component
$\mathcal{H}$.
\begin{equation}
B=e^{-iM_{B}v\cdot x}(\mathcal{N}+\mathcal{H}),
\end{equation}
\begin{equation}
\mathcal{N}=e^{iM_{B}v\cdot x}\frac{1+v\hspace{-0.5em}/}{2}B,~
\mathcal{H}=e^{iM_{B}v\cdot x}\frac{1-v\hspace{-0.5em}/}{2}B,
\end{equation}
where $v_{\mu}=(1,\vec{0})$ is the velocity of the baryon. For the
decuplet baryon, the large component is denoted as
$\mathcal{T}_{\mu}$. Now the decuplet to octet matrix elements of
the electromagnetic current $J_{\mu}$ can be parameterized as
\begin{equation}
<\mathcal N(p)|J_{\mu}|\mathcal T(p^{\prime})>=\bar{u}(p)\mathcal
O_{\rho\mu}(p^{\prime},p)u^{\rho}(p^{\prime}).
\end{equation}
The tensor $\mathcal O_{\rho\mu}$ can be parameterized in terms of
two Lorentz invariant form factors.
\begin{eqnarray}
\begin{split}
\mathcal
O_{\rho\mu}(p^{\prime},p)=\frac{G_{1}}{M_{B}}(q_{\rho}S_{\mu}-q\cdot
Sg_{\rho\mu})+\frac{G_{2}}{4M_{B}^{2}}(q\cdot
vg_{\rho\mu}-q_{\rho}v_{\mu})q\cdot S. \label{eq_newnew_current}
\end{split}
\end{eqnarray}
In the following, we shall use Eq.~(\ref{eq_newnew_current}) to
define the electro quadrupole (E2) and magnetic-dipole (M1)
multipole transtion form factors between the decuplet and octet baryons. The
multipole form factors are
\begin{eqnarray}
G_{M1}  &=&  \frac{2}{3}G_{1}-\frac{\delta}{6M_{T}}G_{1}-\frac{\delta}{12M_{N}}G_{2},\label{eq_formfactor1}\\
G_{E2} & = & \frac{\delta}{6M_{T}}G_{1}-\frac{\delta}{12M_{N}}G_{2}.
\label{eq_formfactor2}
\end{eqnarray}
Accordingly, the $M1$ and $E2$ amplitudes are given by
\begin{eqnarray}
f_{M1} & = & \frac{e}{12M_{B}}(\frac{|q|}{M_{T}M_{B}})^{\frac{1}{2}}\left[(3M_{T}+M_{B})G_{1}-\frac{M_{T}(M_{T}-M_{B})}{2M_{B}}G_{2}\right],\\
f_{E2} & = & -\frac{e}{6M_{B}}\frac{|q|}{M_{T}+M_{B}}(\frac{|q|
M_{T}}{M_{B}})^{\frac{1}{2}}(G_{1}-\frac{M_{T}}{2M_{B}}G_{2}),
\end{eqnarray}
where $|q|=\delta$ in the rest frame of decuplet baryon. The E2 to M1 ratio $R_{\rm EM}$, decay width and transition magnetic moment are expressed as
\begin{eqnarray}
R_{{\rm EM}} & =&\frac{f_{E2}}{f_{M1}}=-\frac{G_{E2}}{G_{M1}},\\
\Gamma(T\rightarrow B\gamma) & = & \frac{\alpha}{16}\frac{(M_{T}^2-M_{B}^2)^3}{M_{T}^3M_{B}^2}(|G_{M1}(q^{2}=0)|^{2}+3|G_{E2}(q^{2}=0)|^{2}),\\
\mu(T\rightarrow
B\gamma)&=&\frac{2M_{T}}{M_{T}+M_{B}}G_{M1}(q^{2}=0)\frac{e}{2M_{B}}.
\end{eqnarray}
where $\alpha=\frac{e^2}{4\pi}=\frac{1}{137}$ is the electromagnetic fine structure
constant.

\section{Chiral Lagrangians}\label{Sec2}

\subsection{The strong interaction chiral Lagrangians}

The pseudoscalar meson fields are introduced as follows,
\begin{equation}
\phi=\left(\begin{array}{ccc}
\pi^{0}+\frac{1}{\sqrt{3}}\eta & \sqrt{2}\pi^{+} & \sqrt{2}K^{+}\\
\sqrt{2}\pi^{-} & -\pi^{0}+\frac{1}{\sqrt{3}}\eta & \sqrt{2}K^{0}\\
\sqrt{2}K^{-} & \sqrt{2}\bar{K}^{0} & -\frac{2}{\sqrt{3}}\eta
\end{array}\right).
\end{equation}
In the framework of ChPT, the chiral connection and axial vector
field are defined as~\cite{Scherer:2002tk,Bernard:1995dp},
\begin{equation}
\Gamma_{\mu}=\frac{1}{2}\left[u^{\dagger}(\partial_{\mu}-ir_{\mu})u+u(\partial_{\mu}-il_{\mu})u^{\dagger}\right],
\end{equation}
\begin{equation}
u_{\mu}\equiv\frac{1}{2}i\left[u^{\dagger}(\partial_{\mu}-ir_{\mu})u-u(\partial_{\mu}-il_{\mu})u^{\dagger}\right],
\end{equation}
where
\begin{equation}
u^{2}=\mathit{U}=\exp(i\phi/f_{0}).
\end{equation}
$f_0$ is the decay constant of the pseudoscalar meson in the chiral
limit. The experimental value of the pion decay constant
$f_{\pi}\approx$ 92.4 MeV while $f_{K}\approx$ 113 MeV,
$f_{\eta}\approx$ 116 MeV.

The lowest order ($\mathcal{O}(p^{2})$) pure meson Lagrangian is
\begin{equation}
\mathcal{L}_{\pi\pi}^{(2)}=\frac{f_{0}^{2}}{4}{\rm
Tr}[\nabla_{\mu}U(\nabla^{\mu}U)^{\dagger}] \label{Eq:meson1},
\end{equation}
where
\begin{equation}
\nabla_{\mu}U=\partial_{\mu}U-ir_{\mu}U+iUl_{\mu}.
\end{equation}
For the electromagnetic interaction,
\begin{equation}
r_{\mu}=l_{\mu}=-eQA_{\mu},Q=\rm{diag}(\frac{2}{3},-\frac{1}{3},-\frac{1}{3}).
\end{equation}
The spin-$\frac{1}{2}$ octet field reads
\begin{equation}
B=\left(\begin{array}{ccc}
\frac{1}{\sqrt{2}}\Sigma^{0}+\frac{1}{\sqrt{6}}\Lambda & \Sigma^{+} & p\\
\Sigma^{-} & -\frac{1}{\sqrt{2}}\Sigma^{0}+\frac{1}{\sqrt{6}}\Lambda & n\\
\Xi^{-} & \Xi^{0} & -\frac{2}{\sqrt{6}}\Lambda
\end{array}\right).
\end{equation}
For the spin-$\frac{3}{2}$ decuplet field, we adopt the
Rarita-Schwinger field $T^{\mu}\equiv
{T^{\mu}}^{abc}$~\cite{Rarita:1941mf}:
\begin{eqnarray}
&&T^{111}=\Delta^{++},T^{112}=\frac{1}{\sqrt{3}}\Delta^{+},T^{122}=\frac{1}{\sqrt{3}}\Delta^{0},T^{222}=\Delta^{-},T^{113}=\frac{1}{\sqrt{3}}\Sigma^{*+},\nonumber\\
&&T^{123}=\frac{1}{\sqrt{6}}\Sigma^{*0},T^{223}=\frac{1}{\sqrt{3}}\Sigma^{*-},T^{133}=\frac{1}{\sqrt{3}}\Xi^{*0},T^{233}=\frac{1}{\sqrt{3}}\Xi^{*-},T^{333}=\Omega^{-}.
\end{eqnarray}
The leading order pseudoscalar meson and baryon interaction
Lagrangians read~\cite{Rarita:1941mf,Jenkins:1992pi}
\begin{eqnarray}
\hat{\mathcal{L}}_{0}^{(1)}&=&{\rm Tr}[\bar{B}(iD\hspace{-0.6em}/-M_{B})B] \nonumber\\
&&+{\rm Tr}{\bar
T^{\mu}[-g_{\mu\nu}(iD\hspace{-0.6em}/-M_{T})+i(\gamma_{\mu}
D_{\mu}+\gamma_{\nu}D_{\mu})-\gamma_{\mu}(iD\hspace{-0.6em}/+M_{T})\gamma_{\nu}]T^{\nu}},
\label{Eq:baryon01}\\
\hat{\mathcal{L}}_{\rm int}^{(1)}&=&\mathcal{C}[{\rm Tr}(\bar
T^{\mu}u_{\mu}B)+{\rm Tr}(\bar B u_{\mu}T^{\mu})]+\mathcal{H}{\rm
Tr}( \bar T^{\mu}g_{\mu\nu}u\hspace{-0.5em}/ \gamma_{5}T^{\nu})
,\label{Eq:baryon02}
\end{eqnarray}
where $M_{B}$ is octet-baryon mass, $M_{T}$ is decuplet-baryon mass,
\begin{eqnarray}
D_{\mu}B&=&\partial_{\mu}B+[\Gamma_{\mu},B], \nonumber\\
D^{\nu}(T^{\mu})_{abc}&=&\partial^{\nu}(T^{\mu})_{abc}+(\Gamma^{\nu})^{d}_{a}(T^{\mu})_{dbc}+(\Gamma^{\nu})^{d}_{b}(T^{\mu})_{adc}+(\Gamma^{\nu})^{d}_{c}(T^{\mu})_{abd}.
\end{eqnarray}
We also need the second order pseudoscalar meson and decuplet-octet
baryon interaction Lagrangians. Recall that
\begin{eqnarray}
8\otimes8 & = & 1\oplus8_{1}\oplus8_{2}\oplus10\oplus\bar{10}\oplus27,\label{Eq:flavor1}\\
8\otimes10 & = & 8\oplus10\oplus27\oplus35.\label{Eq:flavor2}
\end{eqnarray}
Both $u_{\mu}$ and $u_{\nu}$ transform as the adjoint
representation. When the product $u_{\mu} u_{\nu}$ belongs to the
$8_1, 8_2,\bar{10}$ and $27$ flavor representation, we can write
down four independent interaction terms of the second order pseudoscalar meson and baryon Lagrangians:
\begin{eqnarray}
\hat{\mathcal{L}}_{\rm int}^{(2)}&=&\frac{g_{t1}}{4M_{B}} {\rm Tr}(
\bar{B}_{b}^{k}[u_{\mu}, u_{\nu}]_{a}^{i}\epsilon^{jab}
\gamma^{\nu}\gamma_{5}T^{\mu}_{ijk}) +\frac{g_{t2}}{4M_{B}}  {\rm
Tr}(
\bar{B}_{b}^{k}(u_{\mu} u_{\nu})_{al}^{ij}\epsilon^{abl} \gamma^{\nu}\gamma_{5}T^{\mu}_{ijk})\\
&&+\frac{g_{t4}}{4M_{B}} {\rm Tr}( \bar{B}_{b}^{k}\{u_{\nu},
u_{\mu}\}_{a}^{i}\epsilon^{jab} \gamma^{\nu}\gamma_{5}T^{\mu}_{ijk})
+\frac{g_{t3}}{4M_{B}}  {\rm Tr}( \bar{B}_{b}^{l}(u_{\mu}
u_{\nu})_{al}^{ik}\epsilon^{jab}
\gamma^{\nu}\gamma_{5}T^{\mu}_{ijk})+{\rm H.c.}, \label{Eq:baryon03}
\end{eqnarray}
where the superscript denotes the chiral order and $g_{t1,t2,t3,t4}$
is the coupling constant.

In the framework of HBChPT, the baryon field $B$ is decomposed into
the large component $\mathcal{N}$ and the small component
$\mathcal{H}$. We denote the large component of the decuplet baryon
as $\mathcal{T}_{\mu}$. The leading order nonrelativistic
pseudoscalar meson and baryon Lagrangians read~\cite{Jenkins:1992pi}
\begin{equation}
\mathcal{L}_{0}^{(1)}={\rm Tr}[\bar{\mathcal{N}}(iv\cdot
D)\mathcal{N}]-i\bar{\mathcal{T}}^{\mu}(v\cdot
D-\delta)\mathcal{T}_{\mu}, \label{Eq:baryon1}
\end{equation}
\begin{equation}
\mathcal{L}_{\rm
int}^{(1)}=\mathcal{C}(\bar{\mathcal{T}}^{\mu}u_{\mu}\mathcal{N}+\bar{\mathcal{N}}u_{\mu}\mathcal{T}^{\mu})+2\mathcal{H}
\bar{\mathcal{T}}^{\mu}S^{\nu}u_{\nu}\mathcal{T}_{\mu},
\label{Eq:baryon2}
\end{equation}
where $\mathcal{L}_{0}^{(1)}$ and $\mathcal{L}_{\rm int}^{(1)}$ are
the free and interaction parts respectively. $S_{\mu}$ is the
covariant spin-operator. $\delta=M_{T}-M_{B}$ is the octet and
decuplet baryon mass splitting. In the isospin symmetry limit,
$\delta = 0.2937$ GeV. We do not consider the mass difference among
different decuplet baryons. The $\phi\mathcal{N}\mathcal{T}$
coupling $\mathcal{C}=-1.2\pm0.1$ while the
$\phi\mathcal{T}\mathcal{T}$ coupling
$\mathcal{H}=-2.2\pm0.6$~\cite{Butler:1992pn}. For the pseudoscalar
mesons  masses, we use $m_{\pi}=0.140$ GeV, $m_{K}=0.494$ GeV, and
$m_{\eta}=0.550$ GeV. We use the averaged masses for the octet and
decuplet baryons, and $M_{B}=1.158$ GeV, $M_{T}=1.452$ GeV.

The second order pseudoscalar meson and baryon nonrelativistic Lagrangians read
\begin{eqnarray}
\hat{\mathcal{L}}_{\rm int}^{(2)}&=&\frac{g_{t1}}{2M_{B}} {\rm Tr}(
\bar{\mathcal{N}}_{b}^{k}[u_{\mu}, u_{\nu}]_{a}^{i}\epsilon^{jab}
S^{\nu}\mathcal{T}^{\mu}_{ijk}) +\frac{g_{t2}}{2M_{B}}  {\rm Tr}(
\bar{\mathcal{N}}_{b}^{k}(u_{\mu} u_{\nu})_{al}^{ij}\epsilon^{abl}  S^{\nu}\mathcal{T}^{\mu}_{ijk})\\
&&+\frac{g_{t3}}{2M_{B}} {\rm Tr}(
\bar{\mathcal{N}}_{b}^{k}\{u_{\mu}, u_{\nu}\}_{a}^{i}\epsilon^{jab}
S^{\nu}\mathcal{T}^{\mu}_{ijk}) +\frac{g_{t4}}{2M_{B}}  {\rm Tr}(
\bar{\mathcal{N}}_{b}^{l}(u_{\mu} u_{\nu})_{al}^{ik}\epsilon^{jab}
S^{\nu}\mathcal{T}^{\mu}_{ijk})+{\rm H.c.}, \label{Eq:TNUU}
\end{eqnarray}
The above Lagrangians contribute to the decuplet to octet baryon
transition magnetic moments in diagram (d) of
Fig.~\ref{fig:allloop}. After loop integration, the contribution of the $g_{t3}$ term vanishes. Moreover,
the contribution of the
$g_{t4}$ term is exactly proportional to that of the $g_{t2}$ term up to this order.
Thus, there are only two linearly independent low energy constants (LECs)
 $g_{t1}$ and $g_{t2}$  which contribute
to the present investigations of the decuplet to octet baryon transition
form factors up to $\mathcal{O}(p^4)$.
 So we rewrite the second order nonrelativistic
pseudoscalar meson and baryon Lagrangians as
\begin{eqnarray}
\hat{\mathcal{L}}_{\rm int}^{(2)}&=&\frac{\tilde{g}_{t1}}{2M_{B}}
{\rm Tr}( \bar{\mathcal{N}}_{b}^{k}[u_{\mu},
u_{\nu}]_{a}^{i}\epsilon^{jab} S^{\nu}\mathcal{T}^{\mu}_{ijk})
+\frac{\tilde{g}_{t2}}{2M_{B}}  {\rm Tr}(
\bar{\mathcal{N}}_{b}^{k}(u_{\mu} u_{\nu})_{al}^{ij}\epsilon^{abl}
S^{\nu}\mathcal{T}^{\mu}_{ijk})+{\rm H.c.}, \label{Eq:TNUU2}
\end{eqnarray}
where $\tilde{g}_{t1}$ and $\tilde{g}_{t2}$ are the
$\phi\phi\mathcal{T}\mathcal{N}$ coupling constants to be fitted. In
Eq. (28) of Ref.~\cite{Li:2016ezv}, there are two interaction terms
considering  flavor representation. However, only one LEC
contributes to the electromagnetic form factors of the decuplet
baryons.

\subsection{The electromagnetic chiral Lagrangians at $\mathcal{O}(p^{2})$}

The lowest order $\mathcal{O}(p^{2})$ Lagrangian contributes to the
magnetic moments and magnetic dipole form factors of the decuplet
baryons at the tree level~\cite{Jenkins:1992pi}
\begin{equation}
\mathcal{L}_{\mu_{\mathcal T}}^{(2)}=\frac{-ib}{2M_{B}}{\rm
Tr}\bar{\mathcal{T}}^{\mu}F_{\mu\nu}^{+} \mathcal{T}^{\nu},
\label{Eq:baryon3}
\end{equation}
where the coefficient $b$ was extracted in the calculation of the
magnetic moments of the decuplet baryons in Ref.~\cite{Li:2016ezv}.
The chirally covariant QED field strength tensor $F_{\mu\nu}^{\pm}$
is defined as
\begin{eqnarray} \nonumber
F_{\mu\nu}^{\pm} & = & u^{\dagger}F_{\mu\nu}^{R}u\pm
uF_{\mu\nu}^{L}u^{\dagger},\\
F_{\mu\nu}^{R} & = &
\partial_{\mu}r_{\nu}-\partial_{\nu}r_{\mu}-i[r_{\mu},r_{\nu}],\\
F_{\mu\nu}^{L} & = &
\partial_{\mu}l_{\nu}-\partial_{\nu}l_{\mu}-i[l_{\mu},l_{\nu}],
\end{eqnarray}
where $r_{\mu}=l_{\mu}=-eQA_{\mu}$. The operator $F_{\mu\nu}^{\pm}$
transforms as the adjoint representation. Recall that the direct
product $10\otimes\bar{10}=1\oplus8\oplus27\oplus64$ contains only
one adjoint representation. Therefore, there is only one independent
interaction term in the $\mathcal{O}(p^{2})$ Lagrangians for the
magnetic moments of the decuplet baryons.

The lowest order Lagrangians which contribute to the magnetic
moments of the octet baryons at the tree level are,
\begin{equation}
\mathcal{L}_{\mu_{\mathcal N}}^{(2)}=b_F\frac{-i}{4M_{B}}{\rm
Tr}\bar{\mathcal{N}}[S^{\mu},S^{\nu}] [F_{\mu\nu}^{+},
\mathcal{N}]+b_D\frac{-i}{4M_{B}}{\rm
Tr}\bar{\mathcal{N}}[S^{\mu},S^{\nu}]
\{F_{\mu\nu}^{+},\mathcal{N}\}, \label{Eq:baryonoctet}
\end{equation}
where the two LECs were extracted in the calculation of the magnetic
moments of the octet baryons in Ref.~\cite{Meissner:1997hn}:
$b_{D}=3.9$, $b_{F}=3.0$.

The lowest order Lagrangians which contribute to the decuplet-octet
transition magnetic moments at the tree level are
\begin{equation}
\mathcal{L}_{\mu_{\mathcal T\mathcal
N}}^{(2)}=b_2\frac{-i}{2M_{B}}{\rm Tr}\bar{\mathcal{T}}^{\mu}
F_{\mu\nu}^{+}S^{\nu}\mathcal{N} +b_3\frac{-i}{2M_{B}}{\rm
Tr}\bar{\mathcal{T}}^{\mu} F_{\mu\nu}^{+}D^{\nu}\mathcal{N}+{\rm
H.c.},\label{Eq:baryon_trans}
\end{equation}
where $b_2$ is estimated with the help of quark model in
Ref.~\cite{Li:2016ezv}. The $b_3$ term does not contribute to the
transition magnetic moments.

\subsection{The higher order electromagnetic chiral Lagrangians }

The $\mathcal{O}(p^{3})$ Lagrangian which contributes to the electro
quadrupole moments at the tree level reads
\begin{equation}
\mathcal{L}_{\mathbb{Q}_{\mathcal T\mathcal N}}^{(3)}=
c\frac{-1}{4M_{B}^{2}}{\rm
Tr}(\bar{\mathcal{N}}v^{\mu}\partial_{\nu}F_{\rho\mu}^{+}S^{\nu}\mathcal{T}^{\rho})+{\rm
H.c.}. \label{Eq:electro quadrupole}
\end{equation}
To calculate the transition amplitudes to $\mathcal{O}(p^{4})$ and
magnetic moments to $\mathcal{O}(p^{3})$, we also need the
$\mathcal{O}(p^{4})$ electromagnetic chiral Lagrangians at the tree
level. Recalling Eqs.~(\ref{Eq:flavor1}), (\ref{Eq:flavor2}), both
$F_{\mu\nu}^{\pm}$ and $\chi^{+}$ transform as the adjoint
representation. When the product $F_{\mu\nu}^{+} \chi^{+}$ belongs
to the $8_1, 8_2,\bar{10}$ and $27$ flavor representation, we can
write down the chirally invariant $\mathcal{O}(p^{4})$
electromagnetic Lagrangians. Therefore, there exist four independent
interaction terms in the $\mathcal{O}(p^{4})$ chiral Lagrangians.
However, for the $\mathcal{O}(p^{4})$ LEC contribution, we only need
the leading-order terms of the fields $F_{\mu\nu}^{+}$ and
$\chi^{+}$ which are diagonal matrices. Now, only three independent
terms contribute,
\begin{eqnarray}
\mathcal{L}_{\mu_{\mathcal T\mathcal
N}}^{(4)}&=&\hat{d}_{1}\frac{-i}{2M_{B}}{\rm
Tr}(\bar{\mathcal{T}}_{ijk}^{\mu}({F_{\mu\nu}^{+}}{\chi^{+}})_{a}^{i}S^{\nu}{\epsilon^{jab}\mathcal{N}_{b}^{k}})
+\hat{d}_{2}\frac{-i}{2M_{B}}{\rm Tr}(\bar{\mathcal{T}}_{ijk}^{\mu}(\epsilon^{abl}({F_{\mu\nu}^{+}}{\chi^{+}})_{al}^{ij})S^{\nu}\mathcal{N}_{b}^{k})\nonumber\\
 &  & +\hat{d}_{3}\frac{-i}{2M_{B}}{\rm Tr}(\bar{\mathcal{T}}_{ijk}^{\mu}({F_{\mu\nu}^{+}}{\chi^{+}})_{al}^{ik}S^{\nu}{\epsilon^{jab}\mathcal{N}_{b}^{l}})+{\rm H.c.}
. \label{Eq:baryonp40}
\end{eqnarray}
where $\chi^{+}$=diag(0,0,1) at the leading order and the factor
$m_{s}$ has been absorbed in the LECs $\tilde{d}_{1,2,3}$.

There is one more term which contributes to the transition magnetic
moments,
\begin{equation}
\mathcal{L'}_{\mu_{\mathcal T\mathcal
N}}^{(4)}=\tilde{b}^{\prime}\frac{-i}{2M_{B}}{\rm
Tr}(\bar{\mathcal{T}}^{\mu}F_{\mu\nu}^{+}S^{\nu}\mathcal{N}){\rm
Tr}(\chi^{+})+{\rm H.c.}. \label{Eq:baryon_4order}
\end{equation}
However, its contribution can be absorbed through the renomalization
of the LEC $b_2$, i.e.
\begin{eqnarray}
b_2\rightarrow b_2+{\rm Tr}(\chi^{+})\tilde{b}^{\prime}.
\end{eqnarray}

\section{Formalism up to one-loop level}\label{secFormalism}

\begin{figure}[tbh]
\centering
\includegraphics[width=0.9\hsize]{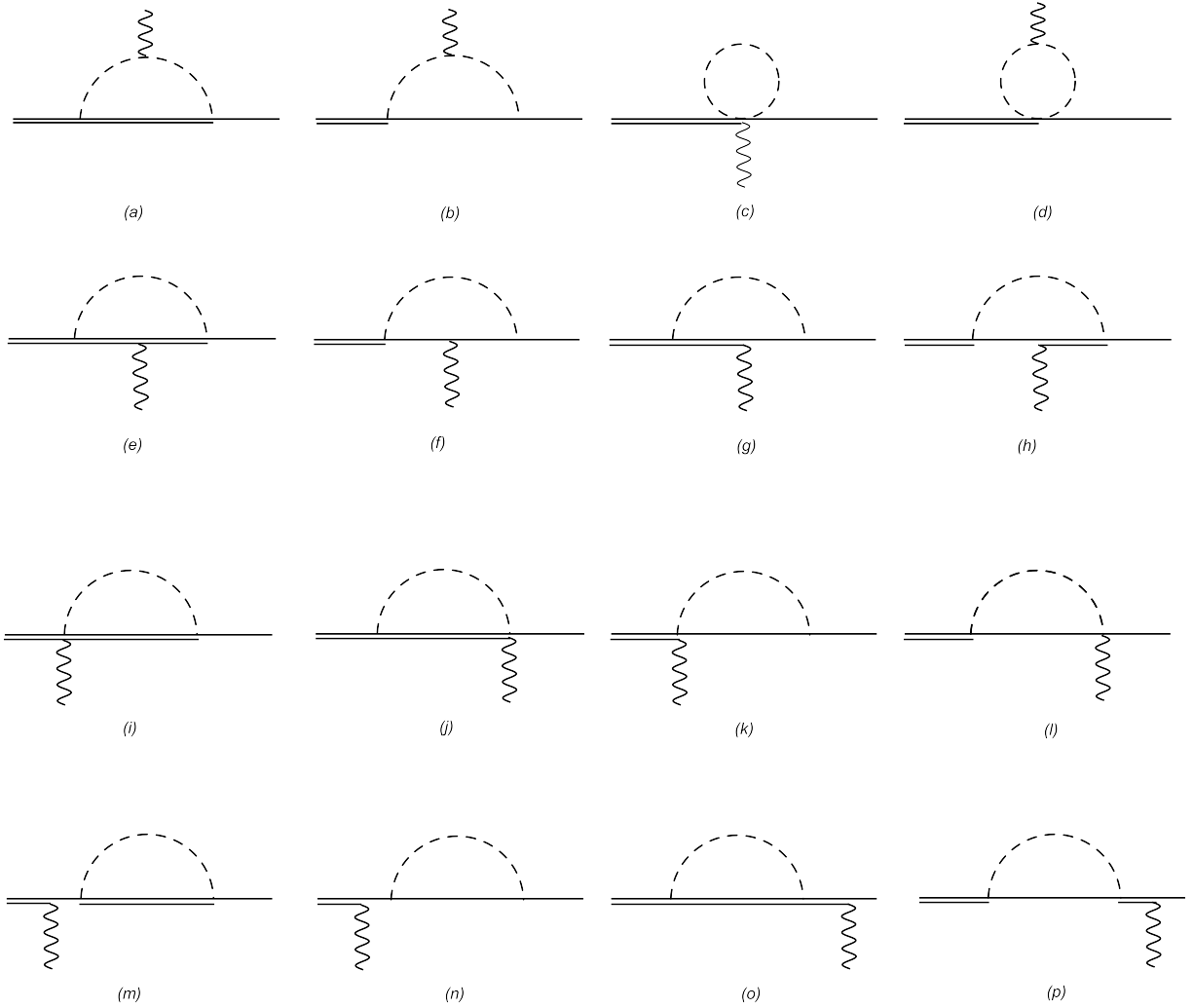}
\caption{The one-loop diagrams where the decuplet (octet) baryon is
denoted by the double (single) solid line. The dashed and wiggly
lines represent the pseudoscalar meson and photon
respectively.}\label{fig:allloop}

\end{figure}

We apply the standard power counting scheme of HBChPT. The chiral
order $D_{\chi}$ of a given diagram is given by~\cite{Ecker:1994gg}
\begin{equation}
D_{\chi}=4N_{L}-2I_{M}-I_{B}+\sum_{n}nN_{n}, \label{Eq:Power
counting}
\end{equation}
where $N_{L}$ is the number of loops, $I_{M}$ is the number of
internal pion lines, $I_{B}$ is the number of internal octet or
decuplet nucleon lines and $N_{n}$ is the number of the vertices from the $n$th
order Lagrangians. As an example, we consider the one-loop diagram
(a) in Fig.~\ref{fig:allloop}. First of all, the number of the
independent loops $N_{L}=1$, the number of the internal pion lines
$I_{M}=2$, the number of the internal octet or decuplet nucleon lines
$I_{B}=1$. For $N_{1}=2$, and $N_{2}=1$ we obtain
$D_{\chi}=4-4-1+2+2=3$.

We use Eq. (\ref{Eq:Power counting}) to count the chiral order
$D_\chi$ of the matrix element of the current, $e \mathcal
O_{\rho\mu}$. We count the unit charge $e$ as $\mathcal O(p^1)$. The
chiral orders of $G_1$ and $G_2$ are $(D_\chi-2)$ and $(D_\chi-3)$,
respectively, since
\begin{equation}
e \mathcal O_{\rho\mu} \sim e p^1 G_1+e p^2 G_2.
\end{equation}
The chiral order of magnetic dipole $G_{M1}$ and electric quadrupole
$G_{E2}$ transition moments are $(D_\chi-1)$ and $(D_\chi-2)$ based
on Eqs. (\ref{eq_formfactor1}) and (\ref{eq_formfactor2}).

\begin{figure}
\centering
\includegraphics[width=0.6\hsize]{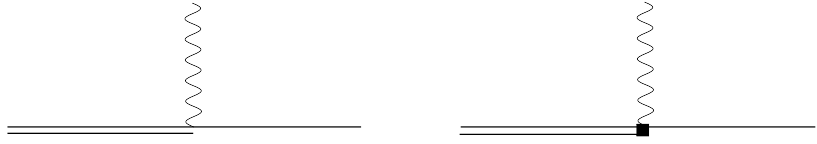}
\caption{The $\mathcal{O}(p^{2})$ and $\mathcal{O}(p^{4})$ tree
level diagram where the decuplet (octet) baryon is denoted by the
double (single) solid line. The left dot and the right black square
represent second- and fourth-order couplings respectively.}
\label{fig:tree}
\end{figure}

Throughout this work, we assume the exact isospin symmetry with
$m_{u}=m_{d}$. The tree-level Lagrangians in Eqs.
~(\ref{Eq:baryon_trans}),(\ref{Eq:baryonp40}) contribute to the
decuplet magnetic moments at $\mathcal{O}(p^{1})$ and
$\mathcal{O}(p^{3})$ as shown in Fig.~\ref{fig:tree}. The
Clebsch-Gordan coefficients for the various decuplet states are
collected in Table~\ref{Magnetic moments}. All decuplet magnetic
moments are given in terms of $\tilde{b}_{2}$, $\tilde{d}_{1}$,
$\tilde{d}_{2}$ and $\tilde{d}_{3}$. $\tilde{b}_{2}$,
$\tilde{c}_{2}$, $\tilde{d}_{1}$, $\tilde{d}_{2}$ and
$\tilde{d}_{3}$ are the liner combinations of LECs $b_{2}$, $c$
$\hat{d}_{1}$, $\hat{d}_{2}$ and $\hat{d}_{3}$. There exist several
interesting relations where we use the baryon "B" to denote
"$T\rightarrow B\gamma$" in the last line,
\begin{eqnarray}
&&{G_{M1}^{\rm tree}}_{\Delta^{+}\rightarrow p\gamma}={G_{M1}^{\rm tree}}_{\Delta^{0}\rightarrow n\gamma}\nonumber\\
&&{G_{M1}^{\rm tree}}_{\Sigma^{*-}\rightarrow\Sigma^{-}\gamma}={G_{M1}^{\rm tree}}_{\Xi^{*-}\rightarrow\Xi^{-}\gamma}\nonumber\\
&&2{G_{M1}^{\rm tree}}_{\Sigma^{*0}\rightarrow\Sigma^{0}\gamma}+{G_{M1}^{\rm tree}}_{\Sigma^{*+}\rightarrow\Sigma^{+}\gamma}={G_{M1}^{\rm tree}}_{\Sigma^{*-}\rightarrow\Sigma^{-}\gamma}\nonumber\\
&&{G_{M1}^{\rm tree}}_{\Sigma^{0}}+\sqrt{3}{G_{M1}^{\rm
tree}}_{\Xi^{0}}={G_{M1}^{\rm tree}}_{p}+(\sqrt{3}-1){G_{M1}^{\rm
tree}}_{\Sigma^{+}}+\sqrt{3}{G_{M1}^{\rm tree}}_{\Lambda}.
\end{eqnarray}

There are sixteen Feynman diagrams at one-loop level as shown in
Fig.~\ref{fig:allloop}. All the vertices in these diagrams come from
Eqs.~(\ref{Eq:meson1}), (\ref{Eq:baryon1}-\ref{Eq:baryon_trans}). In
diagrams (a-b), the meson vertex is from the strong interaction
terms while the photon vertex is from the meson photon interaction
term in Eq.~(\ref{Eq:meson1}). In diagram (c), the
photon-meson-baryon vertex is from the $\mathcal{O}(p^{2})$ tree
level transition magnetic moment interaction in
Eq.~(\ref{Eq:baryon_trans}). In diagram (d), the meson-baryon vertex
is from the second order pseudoscalar meson and baryon Lagrangian in
Eq.~(\ref{Eq:TNUU}) while the photon vertex is also from the meson
photon interaction term. In diagrams (e-h), the meson vertex is from
the strong interaction terms in Eq.~(\ref{Eq:baryon2}) while the
photon vertex from the $\mathcal{O}(p^{2})$ tree level magnetic
moment interaction in Eqs.~(\ref{Eq:baryon3}),
(\ref{Eq:baryonoctet}), (\ref{Eq:baryon_trans}). In diagram (i-l),
the two vertices are from the strong interaction and seagull terms
respectively. In diagrams (m-p), the meson vertex is from the strong
interaction terms while the photon vertex from the
$\mathcal{O}(p^{2})$ tree level transition magnetic moment
interaction in Eq.~(\ref{Eq:baryon_trans}).

The diagrams (a) and (b) contribute to the tensor $e \mathcal
O_{\rho\mu}$ at $\mathcal{O}(p^{3})$ while the diagrams (c-p)
contributes at $\mathcal{O}(p^{4})$. The diagrams (i-l) vanishes in
the heavy baryon mass limit. In particular,
\begin{eqnarray}
 J_{i} & \propto & \bar{u}\int\frac{d^{d}l}{(2\pi)^{d}}q_{\beta}\frac{-iP^{\beta\rho}}{v\cdot l-\delta+i\epsilon}S^{\mu}\frac{i}{l^{2}-m^{2}+i\epsilon}u^{\rho}\protect\\
 \nonumber
 & \propto & v_{\rho}u^{\rho}=0,\\
  J_{j} & \propto & \bar{u}\int\frac{d^{d}l}{(2\pi)^{d}}g_{\beta\mu}\frac{-iP^{\beta\rho}}{v\cdot l-\delta+i\epsilon}S\cdot l\frac{i}{l^{2}-m^{2}+i\epsilon}u^{\rho}\protect\\  \nonumber
 & \propto & S\cdot v=0,\\
   J_{k} & \propto & \bar{u}\int\frac{d^{d}l}{(2\pi)^{d}}S\cdot l\frac{i}{v\cdot l+i\epsilon}g_{\rho\mu}\frac{i}{l^{2}-m^{2}+i\epsilon}u^{\rho}\protect\\  \nonumber
 & \propto & S\cdot v=0,\\
    J_{l} & \propto & \bar{u}\int\frac{d^{d}l}{(2\pi)^{d}}S^{\mu}\frac{i}{v\cdot l+i\epsilon}l^{\rho}\frac{i}{l^{2}-m^{2}+i\epsilon}u^{\rho}\protect\\  \nonumber
 & \propto & v_{\rho}u^{\rho}=0,
 \end{eqnarray}
where $P_{\beta\rho}^{3/2}$ is the non-relativistic spin-$\frac32$
projection operator. In other words, this diagram does not
contribute to the transition magnetic moment in the leading order of
the heavy baryon expansion. The diagrams (m-p) indicate the
corrections from the wave function renormalization.

Summing all the contributions in Fig.~\ref{fig:allloop}, the leading
and next-to-leading order loop corrections to the decuplet to
octet baryon transition magnetic moments can be expressed as
\begin{eqnarray}
\mu_{\mathcal T \mathcal N}^{(2,\rm loop)}& = &\frac{2M_{T}}{M_{T}+M_{B}}\frac{e}{2M_B}[(\frac{2}{3}-\frac{\delta}{6M_{T}})M_{B}\sum_{\phi=\pi,K}\left(\frac{\beta_{\mathcal T}^\phi}{f_\phi^{2}}\mathcal{C}\mathcal{H}a^{M\phi}_{\mathcal N}-\frac{\beta_{\mathcal N}^\phi}{f_\phi^{2}}\mathcal{C}b^{M\phi}_{\mathcal T}\right)\nonumber\\
&&+\frac{\delta M_{B}^2}{3M_{N}}\sum_{\phi=\pi,K}\left(\frac{-\beta_{\mathcal T}^\phi}{3\delta f_\phi^{2}}\mathcal{C}\mathcal{H}a^{E\phi}_{\mathcal N}+\frac{\beta_{\mathcal N}^\phi}{\delta f_\phi^{2}}\mathcal{C}b^{E\phi}_{\mathcal T}\right)],\label{eq:muTN2Loop}\\
\mu_{\mathcal T\mathcal N}^{(3,\rm loop)}& =
&\frac{2M_{T}}{M_{T}+M_{B}}\frac{e}{2M_B}(\frac{2}{3}-\frac{\delta}{6M_{T}})
[\sum_{\phi=\pi,K}(-b_{2}\gamma_{c}^\phi+\frac{1}{2}\gamma_{d}^\phi)\frac{m_\phi^{2}}{8\pi^{2}f_\phi^{2}}\ln\frac{m_\phi}{\mud}
\nonumber\\&& +\sum_{\phi=\pi,K,\eta}(\frac{5}{9\delta
f_{\phi}^{2}}\mathcal{C}\mathcal{H}e^{\phi}_{\mathcal
T}\gamma_{a\mathcal T}^\phi
+\frac{1}{2f_\phi^{2}\delta}\mathcal{C}f^{\phi}_{\mathcal
N}\gamma_{a\mathcal
N}^\phi+\frac{5}{12f_\phi^{2}}\mathcal{H}g^{\phi}_{\mathcal T
\mathcal N}
 b_{2}\gamma^\phi_{a\mathcal T \mathcal N}+\frac{1}{8\delta f_\phi^{2}}\mathcal{C}^{2}h^{\phi}_{ \mathcal N\mathcal T}
 b_{2}\gamma^\phi_{a \mathcal N\mathcal T})]
\nonumber\\&  &
+\frac{2M_{T}}{M_{T}+M_{B}}(\frac{2}{3}-\frac{\delta}{6M_{T}})\sum_{\phi=\pi,K,\eta}[\frac{\mu_{\mathcal
T\mathcal N}^{(1)}}{4f_\phi^{2}}n^{\phi}_{\mathcal
T}\gamma_{f\mathcal N8}^\phi
 +\frac{\mu_{\mathcal T\mathcal N}^{(1)}}{f_\phi^{2}}\frac{-3}{8}\mathcal{C}^{2}m^{\phi}_{\mathcal N}\gamma_{f\mathcal N10}^\phi
 \nonumber\\
&&+\frac{\mu_{\mathcal T\mathcal
N}^{(1)}}{2f_\phi^{2}}\frac{5}{12}\mathcal{H}^{2}m^{\phi}_{\mathcal
N}\gamma_{f\mathcal T10}^\phi+\frac{\mu_{\mathcal T\mathcal
N}^{(1)}}{8f_\phi^{2}}(o^{\pi}_{\mathcal T}+\frac{5}{12}o^{K,\eta
}_{\mathcal T})\mathcal{C}^{2}\gamma_{f\mathcal T8}^\phi],\quad
\label{eq:muTN3Loop}
\end{eqnarray}

\begin{eqnarray}
a^{M\phi}_{\mathcal T} & = &\frac{1}{3456 \pi ^2 \delta }
\begin{cases}\displaystyle
128 \delta ^2-36 \delta ^2 \ln \left(\frac{m_{\phi}^2}{\lambda ^2}\right)-72 \delta  \sqrt{\delta ^2-m_{\phi}^2} \arccosh\left(\frac{\delta }{m_{\phi}}\right)+108 m_{\phi}^2 \arccosh\left(\frac{\delta }{m_{\phi}}\right)^2&\\
+(27 \pi ^2 +24) m_{\phi}^2,  & \phi=\pi,\\ \displaystyle
128 \delta ^2-36 \delta ^2 \ln \left(\frac{m_{\phi}^2}{\lambda ^2}\right)+72 \delta  \sqrt{m_{\phi}^2-\delta ^2} \arccos\left(\frac{\delta }{m_{\phi}}\right)-108 m_{\phi}^2 \arccos\left(\frac{\delta }{m_{\phi}}\right)^2&\\
+(27 \pi ^2+24) m_{\phi}^2, & \phi=K,\eta,
\end{cases}\\
b^{M\phi}_{\mathcal N} & = &\frac{1}{1152 \pi ^2 \delta ^2}
\begin{cases}\displaystyle
32 \delta ^3+24 \pi  m_{\phi}^3-12 \delta ^3 \ln \left(\frac{m_{\phi}^2}{\lambda ^2}\right)+24 i \pi  \left(\delta ^2+2 m_{\phi}^2\right) \sqrt{\delta ^2-m_{\phi}^2}+9 \pi ^2 \delta  m_{\phi}^2+48 \delta  m_{\phi}^2&\\
-24 \sqrt{\delta ^2-m_{\phi}^2} \left(\delta ^2+2 m_{\phi}^2\right)
\arccosh\left(\frac{\delta }{m_{\phi}}\right)-36 \delta  m_{\phi}^2
\arccos\left(-\frac{\delta }{m_{\phi}}\right)^2,  & \phi=\pi,\\
\displaystyle
32 \delta ^3+24 \pi  m_{\phi}^3-12 \delta ^3 \ln \left(\frac{m_{\phi}^2}{\lambda ^2}\right)-24 \sqrt{m_{\phi}^2-\delta ^2} \left(\delta ^2+2 m_{\phi}^2\right) \arccos\left(-\frac{\delta }{m_{\phi}}\right)&\\
+9 \pi ^2 \delta  m_{\phi}^2+48 \delta  m_{\phi}^2-36 \delta
m_{\phi}^2 \arccos\left(-\frac{\delta }{m_{\phi}}\right)^2, &
\phi=K,\eta,
\end{cases}\\
a^{E\phi}_{\mathcal T} & = &\frac{1}{1152 \pi ^2 \delta }
\begin{cases}\displaystyle
-104 \delta ^2+108 \delta ^2 \ln \left(\frac{m_{\phi}^2}{\lambda ^2}\right)+216 \delta  \sqrt{\delta ^2-m_{\phi}^2} \arccosh\left(\frac{\delta }{m_{\phi}}\right)+36 m_{\phi}^2 \arccosh\left(\frac{\delta }{m_{\phi}}\right)^2&\\
+9 \pi ^2 m_{\phi}^2+48 m_{\phi}^2,  & \phi=\pi,\\ \displaystyle
-104 \delta ^2+108 \delta ^2 \ln \left(\frac{m_{\phi}^2}{\lambda
^2}\right)-216 \delta  \sqrt{m_{\phi}^2-\delta ^2}
\arccos\left(\frac{\delta }{m_{\phi}}\right)-36 m_{\phi}^2
\arccos\left(\frac{\delta }{m_{\phi}}\right)^2&\\+9 \pi ^2
m_{\phi}^2+48 m_{\phi}^2, & \phi=K,\eta,
\end{cases}\\
b^{E\phi}_{\mathcal N} & = &\frac{1}{1152 \pi ^2 \delta ^2}
\begin{cases}\displaystyle
-8 \delta ^3+48 \pi  m_{\phi}^3+12 \delta ^3 \ln \left(\frac{m_{\phi}^2}{\lambda ^2}\right)+96 i \pi  m_{\phi}^2 \sqrt{\delta ^2-m_{\phi}^2}-24 i \pi  \delta ^2 \sqrt{\delta ^2-m_{\phi}^2}&\\
-24 \left(4 m_{\phi}^2-\delta ^2\right) \sqrt{\delta ^2-m_{\phi}^2}
\arccosh\left(\frac{\delta }{m_{\phi}}\right)+9 \pi ^2 \delta
m_{\phi}^2+96 \delta  m_{\phi}^2-36 \delta  m_{\phi}^2
\arccos\left(-\frac{\delta }{m_{\phi}}\right)^2,  & \phi=\pi,\\
\displaystyle
-8 \delta ^3+48 \pi  m_{\phi}^3+12 \delta ^3 \ln \left(\frac{m_{\phi}^2}{\lambda ^2}\right)-24 \sqrt{m_{\phi}^2-\delta ^2} \left(4 m_{\phi}^2-\delta ^2\right) \arccos\left(-\frac{\delta }{m_{\phi}}\right)&\\
+9 \pi ^2 \delta  m_{\phi}^2+96 \delta  m_{\phi}^2-36 \delta
m_{\phi}^2 \arccos\left(-\frac{\delta }{m_{\phi}}\right)^2, &
\phi=K,\eta,
\end{cases}\\
e^{\phi}_{\mathcal T} & = &\frac{1}{144 \pi
^2}\begin{cases}\displaystyle \left(6 \delta ^3-9 \delta
m_{\phi}^2\right) \ln \left(\frac{m_{\phi}^2}{\lambda ^2}\right)+12
\left(\delta ^2-m_{\phi}^2\right)^{3/2} \arccosh\left(\frac{\delta
}{m_{\phi}}\right)-2 \left(5 \delta ^3+3 \pi  m_{\phi}^3-6 \delta
m_{\phi}^2\right),  & \phi=\pi,\\ \displaystyle \left(6 \delta ^3-9
\delta  m_{\phi}^2\right) \ln \left(\frac{m_{\phi}^2}{\lambda
^2}\right)+12 \left(m_{\phi}^2-\delta ^2\right)^{3/2}
\arccos\left(\frac{\delta }{m_{\phi}}\right)-2 \left(5 \delta ^3+3
\pi  m_{\phi}^3-6 \delta  m_{\phi}^2\right), & \phi=K,\eta,
\end{cases}\\
f^{\phi}_{\mathcal N} & = &\frac{1}{144 \pi
^2}\begin{cases}\displaystyle
\left(6 \delta ^3-9 \delta  m_{\phi}^2\right) \ln \left(\frac{m_{\phi}^2}{\lambda ^2}\right)+12 \left(\delta ^2-m_{\phi}^2\right)^{3/2} \left(\arccosh\left(\frac{\delta }{m_{\phi}}\right)-i \pi \right)&\\
+2 \left(-5 \delta ^3+3 \pi  m_{\phi}^3+6 \delta  m_{\phi}^2\right),
& \phi=\pi,\\ \displaystyle \left(6 \delta ^3-9 \delta
m_{\phi}^2\right) \ln \left(\frac{m_{\phi}^2}{\lambda ^2}\right)-12
\left(m_{\phi}^2-\delta ^2\right)^{3/2} \arccos\left(-\frac{\delta
}{m_{\phi}}\right)+2 \left(-5 \delta ^3+3 \pi  m_{\phi}^3+6 \delta
m_{\phi}^2\right), & \phi=K,\eta,
\end{cases}\\
g^{\phi}_{\mathcal T\mathcal N} & = & \frac{m_{\phi}^2 }{48 \pi ^2}\left[\ln \left(\frac{m_{\phi}^2}{\lambda ^2}\right)-2\right],\\
h^{\phi}_{\mathcal N\mathcal T} & = &\frac{1}{216 \pi ^2}
\begin{cases}\displaystyle
-14 \delta ^3+\left(6 \delta ^3-9 \delta  m_{\phi}^2\right) \ln \left(\frac{m_{\phi}^2}{\lambda ^2}\right)-6 i \pi  \left(\delta ^2-m_{\phi}^2\right)^{3/2}&\\
+12 \left(\delta ^2-m_{\phi}^2\right)^{3/2}
\arccosh\left(\frac{\delta }{m_{\phi}}\right)+18 \delta  m_{\phi}^2,
& \phi=\pi,\\ \displaystyle
-14 \delta ^3+(6 \delta ^3-9 \delta  m_{\phi}^2 ) \ln \left(\frac{m_{\phi}^2}{\lambda ^2}\right)+6 \left(m_{\phi}^2-\delta ^2\right)^{3/2} \arccos\left(\frac{\delta }{m_{\phi}}\right)&\\
-6 \left(m_{\phi}^2-\delta ^2\right)^{3/2}
\arccos\left(-\frac{\delta }{m_{\phi}}\right)+18 \delta
m_{\phi}^2,& \phi=K,\eta,
\end{cases}\\
n^{\phi}_{\mathcal N}&=&\frac{-1}{16 \pi ^2}
\begin{cases}\displaystyle 2 \delta ^2+\left(m_{\phi}^2-2 \delta
^2\right) \ln \left(\frac{m_{\phi}^2}{\lambda ^2}\right)+4 \delta
\sqrt{\delta ^2-m_{\phi}^2} \arccosh \left(-\frac{\delta
}{m_{\phi}}\right), & \phi=\pi,\\ \displaystyle 2 \delta
^2+\left(m_{\phi}^2-2 \delta ^2\right) \ln
\left(\frac{m_{\phi}^2}{\lambda ^2}\right)-4 \delta
\sqrt{m_{\phi}^2-\delta ^2} \arccos\left(-\frac{\delta
}{m_{\phi}}\right),& \phi=K,\eta,
\end{cases}\\
m^{\phi}_{\mathcal N}&=&\frac{m_{\phi}^2 }{16 \pi ^2}\ln \left(\frac{m_{\phi}^2}{\lambda ^2}\right),\\
o^{\phi}_{\mathcal T}&=&\frac{1}{16 \pi ^2}
\begin{cases}\displaystyle \left(m_{\phi}^2-2 \delta ^2\right) \ln
\left(\frac{m_{\phi}^2}{\lambda ^2}\right)+2 \delta  \left(\delta +2
i \pi  \sqrt{\delta ^2-m_{\phi}^2}\right)-4 \delta  \sqrt{\delta
^2-m_{\phi}^2} \arccosh\left(\frac{\delta }{m_{\phi}}\right), &
\phi=\pi,\\ \displaystyle 2 \delta ^2+\left(m_{\phi}^2-2 \delta
^2\right) \ln \left(\frac{m_{\phi}^2}{\lambda ^2}\right)-4 \delta
\sqrt{m_{\phi}^2-\delta ^2} \arccos\left(-\frac{\delta
}{m_{\phi}}\right),& \phi=K,\eta.
\end{cases}
\end{eqnarray}
where $\mud=1$ GeV is the renormalization scale. The coefficients
$\beta^\phi_{\mathcal T}$ and $\beta^\phi_{\mathcal N}$ arise from
the decuplet and octet intermediate states respectively. We use the
number $n$ within the parenthesis in the superscript of $X^{(n,
...)}$ to indicate the chiral order of $X$.
$\gamma^\phi_{c}$,$\gamma^\phi_{d}$,$\gamma^\phi_{a\mathcal T}$,
$\gamma^\phi_{a\mathcal N}$, $\gamma^\phi_{a\mathcal T\mathcal N}$,
$\gamma^\phi_{f\mathcal N8}$, $\gamma^\phi_{f\mathcal N10}$,
$\gamma^\phi_{f\mathcal T10}$ and $\gamma^\phi_{f\mathcal T8}$ arise
from the corresponding diagrams in Fig.~\ref{fig:allloop}. We
collect their explicit expressions in Tables~\ref{table:beta},
\ref{table:cd}, \ref{table:ef}, \ref{table:gh}, \ref{table:mn},
\ref{table:op} in the Appendix \ref{appendix-B}.

With the low energy counter terms and loop contributions
(\ref{eq:muTN2Loop}, \ref{eq:muTN3Loop}), we obtain the magnetic
moments,
\begin{equation}
\mu_{\mathcal T\mathcal N}=\left\{\mu_{\mathcal T\mathcal
N}^{(1)}\right\}+\left\{\mu_{\mathcal T\mathcal N}^{(2,\rm
loop)}\right\}+\left\{\mu_{\mathcal T\mathcal N}^{(3,\rm
tree)}+\mu_{\mathcal T\mathcal N}^{(3,\rm loop)}\right\}
\end{equation}
where $\mu_{\mathcal T}^{(1)}$ and $\mu_{\mathcal T}^{(3,\rm tree)}$
are the tree-level magnetic moments from
Eqs.~(\ref{Eq:baryon_trans}),(\ref{Eq:baryonp40}).

Summing all the contributions to electric quadrupole moments in
Fig.~\ref{fig:allloop}, the leading and next-to-leading order loop
corrections can be expressed as
\begin{eqnarray}
G_{E2}^{(1,\rm loop)}& = &\frac{\delta}{6M_{T}}M_{B}\sum_{\phi=\pi,K}\left(\frac{\beta_{\mathcal T}^\phi}{f_\phi^{2}}\mathcal{C}\mathcal{H}a^{M\phi}_{\mathcal N}-\frac{\beta_{\mathcal N}^\phi}{f_\phi^{2}}\mathcal{C}b^{M\phi}_{\mathcal T}\right)+\frac{\delta M_{B}^2}{3M_{N}}\sum_{\phi=\pi,K}\left(\frac{-\beta_{\mathcal T}^\phi}{3\delta f_\phi^{2}}\mathcal{C}\mathcal{H}a^{E\phi}_{\mathcal N}+\frac{\beta_{\mathcal N}^\phi}{\delta f_\phi^{2}}\mathcal{C}b^{E\phi}_{\mathcal T}\right), \label{eq:eleqTN1Loop}\\
G_{E2}^{(2,\rm loop)}& = &0. \label{eq:eleqTN2Loop}
\end{eqnarray}

From the tensor $e \mathcal O_{\rho\mu\sigma}$ up to
$\mathcal{O}(p^{4})$, with the low energy counter terms and loop
contributions (\ref{eq:eleqTN1Loop}, \ref{eq:eleqTN2Loop}), we
obtain the electric quadrupole moments at the next-to-leading order,
\begin{equation}
G_{E2}=\left\{G_{E2}^{(1,\rm tree)}+G_{E2}^{(1,\rm
loop)}\right\}+\left\{G_{E2}^{(2,\rm loop)}\right\}
\end{equation}
where $G_{E2}^{(1,\rm tree)}$ is the tree-level  electro quadrupole
moments from Eq.~(\ref{Eq:electro quadrupole}).

\section{NUMERICAL RESULTS AND DISCUSSIONS}\label{Sec6}

\begin{table}
  \centering
\begin{tabular}{c|ccccccc}
\toprule[1pt]\toprule[1pt] Process($G_{\rm M1}$) &
$\mathcal{O}(p^{1})$  & $\mathcal{O}(p^{2})$ & $\mathcal{O}(p^{3})$
tree & $\mathcal{O}(p^{3})$ loop & Total Fit A  & Total Fit B  &
PDG\tabularnewline \midrule[1pt] $\Delta^{+}\rightarrow p\gamma$
& $-\frac{2}{\sqrt{3}}\tilde{b}_{2}$ & 1.99 & $0$ &
$1.83+0.10\tilde{g}_{t1}+0.04\tilde{g}_{t2}$ & -3.18 & -3.22&
-3.12(14)\tabularnewline

$\Delta^{0}\rightarrow n\gamma$ &
$-\frac{2}{\sqrt{3}}\tilde{b}_{2}$ & 1.99 & $0$ &
$1.83+0.10\tilde{g}_{t1}+0.04\tilde{g}_{t2}$ & -3.18 & -3.22&
-3.12(14)\tabularnewline

$\Sigma^{*+}\rightarrow\Sigma^{+}\gamma$ &
$\frac{2}{\sqrt{3}}\tilde{b}_{2}$ & -3.15 &
$\frac{2}{3\sqrt{3}}\tilde{d}_{1}+\frac{4\sqrt{3}}{9}\tilde{d}_{2}$
& $-1.46-0.15\tilde{g}_{t1}-0.12\tilde{g}_{t2}$ & 4.05 & 4.03&
4.05(49)\tabularnewline

$\Sigma^{*0}\rightarrow\Sigma^{0}\gamma$ &
$-\frac{1}{\sqrt{3}}\tilde{b}_{2}$ & 1.72 &
$-\frac{2}{3\sqrt{3}}\tilde{d}_{1}-\frac{\sqrt{3}}{9}\tilde{d}_{2}$
& $0.72+0.09\tilde{g}_{t1}+0.06\tilde{g}_{t2}$ & -2.12 &
-2.11&\textemdash{} \tabularnewline

$\Sigma^{*0}\rightarrow\Lambda\gamma$ & $\tilde{b}_{2}$ & -1.98
& $\frac{1}{3}\tilde{d}_{2}+\frac{2\sqrt{3}}{3}\tilde{d}_{3}$ &
$-0.49-0.09\tilde{g}_{t1}-0.10\tilde{g}_{t2}$ & 3.29& 3.22 &
3.25(46)\tabularnewline

$\Sigma^{*-}\rightarrow \Sigma^{-}\gamma$ & 0 & 0.29 &
$-\frac{2}{3\sqrt{3}}\tilde{d}_{1}+\frac{2\sqrt{3}}{9}\tilde{d}_{2}$
& $-0.01+0.04\tilde{g}_{t1}$ & -0.19& -0.20 &
$<0.78(04)$\tabularnewline

$\Xi^{*0}\rightarrow\Xi^{0}\gamma$ &
$\frac{2}{\sqrt{3}}\tilde{b}_{2}$ & -3.15 &
$\frac{2}{3\sqrt{3}}\tilde{d}_{1}+\frac{4\sqrt{3}}{9}\tilde{d}_{2}+\frac{2\sqrt{3}}{3}\tilde{d}_{3}$
& $0.49-0.15\tilde{g}_{t1}-0.12\tilde{g}_{t2}$ & 4.89& 4.10 &
$<4.90(53)$\tabularnewline

$\Xi^{*-}\rightarrow \Xi^{-}\gamma$ & 0 & 0.29 &
$-\frac{2}{3\sqrt{3}}\tilde{d}_{1}+\frac{2\sqrt{3}}{9}\tilde{d}_{2}$
& $0.36+0.04\tilde{g}_{t1}$ & 0.18 & 0.18&
$<4.90(53)$\tabularnewline \bottomrule[1pt]\bottomrule[1pt]
\end{tabular}
\caption{The decuplet to octet baryon transition $G_{M1}(q^2=0)$ to the
next-to-next-to-leading order (in unit of 1).} \label{Magnetic
moments}
\end{table}

\begin{table}
  \centering
\begin{tabular}{c|ccccc}
\toprule[1pt]\toprule[1pt] Process($\mu_{\mathcal T \mathcal N}$) &
$\mathcal{O}(p^{1})$  & $\mathcal{O}(p^{2})$ & $\mathcal{O}(p^{3})$
Fit A &$\mathcal{O}(p^{3})$ Fit B &  PDG\tabularnewline
\midrule[1pt] $\Delta^{+}\rightarrow p\gamma$ & -3.43(15) &
-3.43(65) & -3.50(67) & -3.54(68)& -3.43(15)\tabularnewline

$\Delta^{0}\rightarrow n\gamma$ & -3.43(15) & -3.43(65) &
-3.50(67) &-3.54(68)& -3.43(15) \tabularnewline

$\Sigma^{*+}\rightarrow\Sigma^{+}\gamma$ & 3.43(15) & 2.16(87) &
4.46(94) &4.43(93)& 4.45(54)\tabularnewline

$\Sigma^{*0}\rightarrow\Sigma^{0}\gamma$ & -1.72(08) & -0.92(46)
& -2.34(50)&-2.32(49) &\textemdash{}\tabularnewline

$\Sigma^{*0}\rightarrow\Lambda\gamma$ & 2.97(13) &2.69(61) &
3.62(63)&3.54(62) & 3.69(50)\tabularnewline

$\Sigma^{*-}\rightarrow \Sigma^{-}\gamma$ & 0 & 0.32(06) &
-0.21(07) &-0.22(07)& $<0.85(05)$\tabularnewline

$\Xi^{*0}\rightarrow\Xi^{0}\gamma$ & 3.43(15) & 2.16(87) &
5.38(96) & 4.51(80)& $<5.39(58)$\tabularnewline

$\Xi^{*-}\rightarrow \Xi^{-}\gamma$ & 0 & 0.32(06) & 0.20(06) &
0.19(06)& $<5.39(58)$\tabularnewline
\bottomrule[1pt]\bottomrule[1pt]
\end{tabular}
\caption{The decuplet to octet baryon transition magnetic moments when the
chiral expansion is truncated at $\mathcal{O}(p^{1})$,
$\mathcal{O}(p^{2})$, and $\mathcal{O}(p^{3})$, respectively (in
unit of $\mu_{N}$).} \label{various orders Magnetic moments}
\end{table}

We collect our numerical results of the baryon decuplet to octet baryon
transition magnetic moments to the next-to-next-to-leading order in
Table~\ref{Magnetic moments}. We also compare the numerical results
of the transition magnetic moments when the chiral expansion is
truncated at $\mathcal{O}(p^{1})$, $\mathcal{O}(p^{2})$ and
$\mathcal{O}(p^{3})$ respectively in Table~\ref{various orders
Magnetic moments}.

At the leading order $\mathcal{O}(p^{1})$, there is only one unknown
low energy constant $\tilde{b}_2$. We use the precise experimental
measurement of the $\Delta\rightarrow N\gamma$ transition
magnetic moment $\mu_{\Delta\rightarrow
N\gamma}=(-3.43\pm0.15)\mu_{N}$ as input to extract
$\tilde{b}_2=2.97\pm0.13$. The magnetic moments of the other
decuplet baryons are given in the second column in
Table~\ref{various orders Magnetic moments}. Notice that the
$\mathcal{O}(p^{1})$ tree level transitions
$\Delta^{+}\rightarrow p\gamma$ and $\Delta^{0}\rightarrow
n\gamma$ are the same. In fact, this equation holds to every order because of the exact SU(2) spin-flavor symmetry. The $\mathcal{O}(p^{1})$ tree level
transitions $\Sigma^{*-}\rightarrow \Sigma^{-}\gamma$ and
$\Xi^{*-}\rightarrow \Xi^{-}\gamma$ are zero because of the
famous U-spin symmetry as can be seen from Table~\ref{table:quarkmodel} in the Appendix~\ref{appendix-D}.

Up to $\mathcal{O}(p^{2})$, we need include both the leading
tree-level magnetic moments and the $\mathcal{O}(p^{2})$ loop
corrections. At this order, all the coupling constants are
well-known. There do not exist new LECs. Again, we use the
experimental value of the $\Delta\rightarrow N\gamma$ transition
magnetic moment $\mu_{\Delta\rightarrow
N\gamma}=(-3.43\pm0.15)\mu_{N}$ as input to extract the LEC
$\tilde{b}_2=4.87\pm0.13$. We list the numerical results in the
third column in Table~\ref{various orders Magnetic moments}, where
the errors in the brackets are dominated by the errors of the
coupling constants $\mathcal{C}, \mathcal{H}$ in Eq.
(\ref{Eq:baryon2}). Notice that the $\Sigma^{*+}\rightarrow
\Sigma^{+}\gamma$ transition magnetic moment to $\mathcal{O}(p^{2})$
is quite small compared to the experimental value, which will be
improved when the $\mathcal{O}(p^{3})$ contribution is included. In
other approaches in Table~\ref{Comparison of magnetic moments}, the
transition magnetic moment of $\Sigma^{*+}\rightarrow
\Sigma^{+}\gamma$ is also smaller than that of
$\Delta\rightarrow N\gamma$.

Up to $\mathcal{O}(p^{3})$, there are six unknown LECs:
$\tilde{b}_{2}$, $\tilde{g}_{t1,t2}$, $\tilde{d}_{1,2,3}$. Two
schemes will be introduced to fit all LECs. In the first fit (Fit
A), we use the experimental values of the transition magnetic
moments of $\Delta\rightarrow N\gamma$,
$\Sigma^{*+}\rightarrow\Sigma^{+}\gamma$,
$\Sigma^{*0}\rightarrow\Lambda\gamma$, the upper limit of the
transition magnetic moment of $\Xi^{*0}\rightarrow\Xi^{0}\gamma$
and $\mu_{\Sigma^{*-}\rightarrow \Sigma^{-}\gamma}=0$,
$\mu_{\Xi^{*-}\rightarrow \Xi^{-}\gamma}=0$ to extract the six
LECs: $\tilde{b}_{2}=5.26$, $\tilde{g}_{t1}=-3.73$,
$\tilde{g}_{t2}=-12.91$, $\tilde{d}_{1}=0.93$, $\tilde{d}_{2}=0.10$,
$\tilde{d}_{3}=-0.96$. We list the numerical results up to
$\mathcal{O}(p^{3})$ in the fourth column in Table~\ref{various
orders Magnetic moments} .

In order to study the convergence of the chiral expansion, we show
the numerical results at each order for the transition magnetic
moments:
\begin{eqnarray}
\mu_{\Delta^{+}\rightarrow p\gamma}&=&-6.68\times(1 - 0.33 - 0.15)=-3.50,\nonumber\\
\mu_{\Delta^{0}\rightarrow n\gamma}&=&-6.68\times(1 - 0.33 - 0.15)=-3.50,\nonumber\\
\mu_{\Sigma^{*+}\rightarrow \Sigma^{+}\gamma}&=&6.68\times (1 - 0.52 + 0.19)=4.46,\nonumber\\
\mu_{\Sigma^{*0}\rightarrow \Sigma^{0}\gamma}&=&-3.34\times (1 - 0.57 + 0.27)=-2.34,\nonumber\\
\mu_{\Sigma^{*0}\rightarrow \Lambda\gamma}&=&5.78\times (1 - 0.38 + 0.01)=3.62,\nonumber\\
\mu_{\Sigma^{*-}\rightarrow \Sigma^{-}\gamma}&=&0.32 \times(0+1 - 1.65)=-0.21,\nonumber\\
\mu_{\Xi^{*0}\rightarrow \Xi^{0}\gamma}&=&6.68 \times(1 - 0.52 + 0.32)=5.38,\nonumber\\
\mu_{\Xi^{*-}\rightarrow \Xi^{-}\gamma}&=&0.32\times (0+1 -
0.38)=0.20.
\end{eqnarray}
For the U-spin forbidden processes, their magnetic moments vanish at
$\mathcal{O}(p^{1})$. Their total magnetic moments arise from the
loop contributions at $\mathcal{O}(p^{2,3})$ and the tree-level LECs
$d_{1,2,3}$ at $\mathcal{O}(p^{3})$ which are related to the strange
quark mass correction. For the other processes, one observes rather
good convergence of the chiral expansion and the leading order term
dominates in these channels.

In the second fit (Fit B), considering transition magnetic moments
of $\Sigma^{*+}\rightarrow\Sigma^{+}\gamma$ and
$\Xi^{*0}\rightarrow\Xi^{0}\gamma$ are the same in quark model
as shown in Table~\ref{table:quarkmodel}, we use the experimental
value of the transition magnetic moments of $\Delta\rightarrow
N\gamma$, $\Sigma^{*+}\rightarrow\Sigma^{+}\gamma$,
$\Sigma^{*0}\rightarrow\Lambda\gamma$ and
$\mu_{\Xi^{*0}\rightarrow\Xi^{0}\gamma}=\mu_{\Sigma^{*+}\rightarrow\Sigma^{+}\gamma}$,
$\mu_{\Sigma^{*-}\rightarrow \Sigma^{-}\gamma}=0$,
$\mu_{\Xi^{*-}\rightarrow \Xi^{-}\gamma}=0$ to extract the six
LECs: $\tilde{b}_{2}=4.22$, $\tilde{g}_{t1}=-8.18$,
$\tilde{g}_{t2}=-31.69$, $\tilde{d}_{1}=-0.96$,
$\tilde{d}_{2}=-0.32$, $\tilde{d}_{3}=-1.62$. We list the numerical
results up to $\mathcal{O}(p^{3})$ in the fifth column in
Table~\ref{various orders Magnetic moments}. We also show the
numerical results at each order in Fit B:
\begin{eqnarray}
\mu_{\Delta^{+}\rightarrow p\gamma}&=&-5.36 \times(1 - 0.41 + 0.07)=-3.54,\nonumber\\
\mu_{\Delta^{0}\rightarrow n\gamma}&=&-5.36\times (1 - 0.41 + 0.07)=-3.54,\nonumber\\
\mu_{\Sigma^{*+}\rightarrow \Sigma^{+}\gamma}&=&5.36\times (1 - 0.65 + 0.47)=4.43,\nonumber\\
\mu_{\Sigma^{*0}\rightarrow \Sigma^{0}\gamma}&=&-2.68\times (1 - 0.70 + 0.57)=-2.32,\nonumber\\
\mu_{\Sigma^{*0}\rightarrow \Lambda\gamma}&=&4.64\times (1 - 0.47 + 0.23)=3.54,\nonumber\\
\mu_{\Sigma^{*-}\rightarrow \Sigma^{-}\gamma}&=&0.32\times (0+1 - 1.69)=-0.22,\nonumber\\
\mu_{\Xi^{*0}\rightarrow \Xi^{0}\gamma}&=&5.36\times (1 - 0.65 + 0.49)=4.51,\nonumber\\
\mu_{\Xi^{*-}\rightarrow \Xi^{-}\gamma}&=&0.32 \times(0+1 -
0.40)=0.19.
\end{eqnarray}

We collect our numerical results of the decuplet to octet baryon
transition electro quadrupole moments to next-to-leading order in
Table~\ref{table:Electro quadrupole}. Up to $\mathcal{O}(p^{2})$, we
need include both the leading tree-level magnetic moments and the
$\mathcal{O}(p^{1,2})$ loop corrections. The $\mathcal{O}(p^{2})$
loop corrections in Fig~\ref{fig:allloop} are zero, so there is only
one unknown low energy constant $\tilde{c}$ from
Eq.~(\ref{Eq:electro quadrupole}). We use the experimental value of
the $\Delta\rightarrow N\gamma$ transition  magnetic dipole
moment $G_{\rm M1}=-3.12\pm0.14$ and the E2 to M1 ratio $R_{\rm EM}=(-2.5\pm0.5)\%$~\cite{Patrignani} as input to extract the LEC
$\tilde{c}_2=0.475\pm0.143$. We list the numerical results of the
transition electro quadrupole moments and the E2 to M1 ratio $R_{\rm EM}$ in
Table~\ref{table:Electro quadrupole}.

We also calculate the M1 and E2 amplitudes and decay width of the
decuplet to octet baryon transitions in the
Appendix~\ref{appendix-C}. Both fits A and B lead to the same decay
width for $\Sigma^{*0}\rightarrow\Sigma^{0}\gamma$. The E2 amplitude of
the $\Sigma^{*0}\rightarrow\Sigma^{0}\gamma$ channel does not have
any imaginary part because the $\pi^+$ and $\pi^-$ loop
contributions cancel each other as shown in Table~\ref{table:beta}.
In other words, the pion loop contributions with the intermediate
baryons $\Sigma^{*+}$ and $\Sigma^{*-}$, $\Sigma^{+}$ and
$\Sigma^{-}$ cancel each other due to the exact SU(2) flavor
symmetry. The extracted M1 and E2 transition amplitudes and
radiative decay widths may be useful for future experimental
measurement.

\begin{table}
\centering
\begin{tabular}{c|cccc}
\toprule[1pt]\toprule[1pt] Process($G_{\rm E2}$) & Tree value& Loop
value& Total value & $R_{\rm EM}$\tabularnewline \midrule[1pt]
$\Delta^{+}\rightarrow p\gamma$ & -0.548 & 0.473 & -0.075(26) &
-2.5(9)\%\tabularnewline

$\Delta^{0}\rightarrow n\gamma$ & -0.548 & 0.473 & -0.075(26) &
-2.5(9)\%\tabularnewline

$\Sigma^{*+}\rightarrow\Sigma^{+}\gamma$ & 0.548 & -0.502 &
0.046(13) &1.1(3)\% \tabularnewline

$\Sigma^{*0}\rightarrow\Sigma^{0}\gamma$ & -0.274 & 0.255 &
-0.019(05) & 0.9(3)\%\tabularnewline

$\Sigma^{*0}\rightarrow\Lambda\gamma$ & 0.475 & -0.416 & 0.059(19) &
1.8(6)\%\tabularnewline

$\Sigma^{*-}\rightarrow \Sigma^{-}\gamma$ & 0 & 0.007 & 0.007(4) &
-3.7(2.1)\%\tabularnewline

$\Xi^{*0}\rightarrow\Xi^{0}\gamma$ & 0.548 & -0.502 & 0.046(16)
&0.9(3)\% \tabularnewline

$\Xi^{*-}\rightarrow \Xi^{-}\gamma$ & 0 & 0.007 & 0.007(4) &
3.9(2.2)\%\tabularnewline
 \bottomrule[1pt]\bottomrule[1pt]
\end{tabular}
 \caption{Electro quadrupole transition moments(in  unit of $1$).}
\label{table:Electro quadrupole}
\end{table}

\section{Conclusions}\label{Sec7}

In short summary, we have systematically studied the decuplet to
octet baryon transition magnetic moments up to the
next-to-next-to-leading order in the framework of the heavy baryon
chiral perturbation theory. With both the octet and decuplet baryon
intermediate states in the chiral loops, we have systematically
calculated the chiral corrections to the transition magnetic moments
order by order. The chiral expansion converges rather well for the
charged channels. In Table~\ref{Comparison of magnetic moments}, we
compare our results obtained in the HBChPT with those from other
model calculations such as lattice QCD
(LQCD)~\cite{Leinweber:1992pv}, chiral quark model
(ChQM)~\cite{Kim:2005gz}, relativistic quark model
(RQM)~\cite{Ramalho:2013uza}, effective mass quark model
(EQM)~\cite{Dhir:2009ax}, meson cloud (MS)~\cite{Ramalho:2013iaa},
U-spin~\cite{Keller:2013hza}, QCD sum rules
(QCD-SR)~\cite{Wang:2009bh,Hong:2007pr} and large
$N_{c}$~\cite{Jenkins:2011dr,Lebed:2004fj}. We also list the
experimental values from the PDG~\cite{Patrignani}. One may observe
the qualitatively similar features for the transition magnetic
moments. We have also systematically calculated the electro
quadrupole moments to next-to-leading order and obtained the E2 to
M1 ratio $R_{\rm EM}$ for decuplet to octet baryon transition, which
suggests the d-wave component and deformed structure of the octet
and decuplet baryons. Our results may be useful for future
experimental measurement of the electro quadrupole multipole
moments.

The decuplet to octet baryon transition magnetic moments of
$\Delta^{+}\rightarrow p\gamma$ and $\Delta^{0}\rightarrow
n\gamma$ are always the same because of the exact SU(2) spin-flavor
symmetry. Comparing the $\mathcal{O}(p^{2})$ and
$\mathcal{O}(p^{3})$ $\Sigma^{*+}\rightarrow\Sigma^{+}\gamma$
transition magnetic moments to the experimental values, we want to
emphasize the importance of the next-to-next-to-leading order chiral
correction. As the current experimental data is not enough, we
introduce two schemes to fit all LECs. Both fitting schemes lead to
reasonably good convergence of the chiral expansion and agreement
with the experimental data. We hope that more decuplet to octet
baryon transition magnetic moments like
$\Xi^{*0}\rightarrow\Xi^{0}\gamma$ will be measured more
precisely in future experiments. Moreover, the analytical
expressions derived in this work may be useful to the possible
chiral extrapolation of the lattice simulations of the decuplet to
octet baryon transition electromagnetic properties in the coming
future.

 \begin{table}
  \centering
\begin{tabular}{c|ccccccc}
\toprule[1pt]\toprule[1pt] Process & $\Delta\rightarrow N\gamma$ &
$\Sigma^{*+}\rightarrow\Sigma^{+}\gamma$ &
$\Sigma^{*0}\rightarrow\Sigma^{0}\gamma$ &
$\Sigma^{*0}\rightarrow\Lambda\gamma$ &
$\Sigma^{*-}\rightarrow\Sigma^{-}\gamma$ &
$\Xi^{*0}\rightarrow\Xi^{0}\gamma$ &
$\Xi^{*-}\rightarrow\Xi^{-}\gamma$\tabularnewline \midrule[1pt]
LQCD~\cite{Leinweber:1992pv} & 2.46 & 2.61 & 1.07 & \textemdash{} &
-0.47 & -2.77 & 0.47\tabularnewline

ChQM~\cite{Kim:2005gz} & -3.31 & 2.17 & \textemdash{} & -2.74 &
-0.59 & 2.23 & -0.59\tabularnewline

RQM~\cite{Ramalho:2013uza} & 3.25 & 2.59 & 1.07 & 2.86 & -0.46 &
2.71 & -0.47\tabularnewline

EQM~\cite{Dhir:2009ax} & 2.63 & 2.33 & 1.02 & 2.28 & 0.30 & 2.33 &
0.30\tabularnewline

MS~\cite{Ramalho:2013iaa} & 3.32 & 3.54 & 1.61 & 3.39 & -0.34 & 3.62
& -0.42\tabularnewline

U-spin~\cite{Keller:2013hza} & \textemdash{} & 3.22 & 1.61 & 2.68 &
0 & 3.21 &\textemdash{} \tabularnewline

QCD-SR~\cite{Wang:2009bh} & 3.86 & 3.38 & 1.47 & 4.44 & -0.57 &
-1.24 & 0.23\tabularnewline

QCD-SR~\cite{Hong:2007pr} & -2.76 & 2.24 & 1.01 & -2.46 & -0.22 &
2.46 & -0.27\tabularnewline

Large $Nc$~\cite{Jenkins:2011dr} & 3.51 & 2.96 & 1.34 & 2.96 & -0.27
& 2.96 &\textemdash{} \tabularnewline

Large $Nc$~\cite{Lebed:2004fj} & 3.51 & 2.97 & 1.39 & 2.93 & -0.19 &
2.96 & -0.19\tabularnewline

This work (fit A)& -3.50 & 4.46 & -2.34 & 3.62 & -0.21 & 5.38 &
0.20\tabularnewline

This work (fit B)& -3.54 & 4.43 & -2.34 & 3.54 & -0.22 & 4.51 &
0.19\tabularnewline

PDG/$\mu_{N}$ & -3.43(15) & 4.45(54) &\textemdash{}  & 3.69(50) &
$<0.85(05)$ & $<5.39(58)$ & $<5.39(58)$\tabularnewline
\bottomrule[1pt]\bottomrule[1pt]
\end{tabular}
\caption{Comparison of the decuplet to octet baryon transition magnetic
moments in literature including lattice
QCD (LQCD)~\cite{Leinweber:1992pv}, chiral quark
model (ChQM)~\cite{Kim:2005gz}, relativistic quark
model (RQM)~\cite{Ramalho:2013uza}, effective mass quark
model (EQM)~\cite{Dhir:2009ax}, meson
cloud (MS)~\cite{Ramalho:2013iaa}, U-spin~\cite{Keller:2013hza}, QCD
sum rules (QCD-SR)~\cite{Wang:2009bh,Hong:2007pr}, large
$N_{c}$~\cite{Jenkins:2011dr,Lebed:2004fj}, and
PDG~\cite{Patrignani}(in unit of $\mu_{N}$).}
  \label{Comparison of magnetic moments}
 \end{table}

\section*{ACKNOWLEDGMENTS}

H. S. Li is very grateful to B. Zhou and L. Meng for very helpful
discussions. This project is supported by the National Natural
Science Foundation of China under Grants 11575008, 11621131001 and
973 program.

\begin{appendix}

\section{Integrals and loop functions} \label{appendix-A}

We collect some common integrals and loop functions in this
appendix.

\subsection{Integrals with one or two meson propagators}
\begin{eqnarray}
\Delta & = & i\int\frac{d^{d}l \,\mud^{4-d}}{(2\pi)^{d}}\frac{1}{l^{2}-m^{2}+i\epsilon}= 2m^{2}(L(\mud)+\frac{1}{32\pi^{2}}\ln\frac{m^{2}}{\mud^{2}}),\\
L(\mud) & = &
\frac{\mud^{d-4}}{16\pi^{2}}[\frac{1}{d-4}-\frac{1}{2}(\ln4\pi+1+\Gamma^{\prime}(1))].
\end{eqnarray}
\begin{eqnarray}
I_{0}(q^{2})&=&i\int\frac{d^{d}l\,\mud^{4-d}}{(2\pi)^{d}}\frac{1}{(l^{2}-m^{2}+i\epsilon)((l+q)^{2}-m^{2}+i\epsilon)}\nonumber\\
&=&\begin{cases} \displaystyle
-\frac{1}{16\pi^{2}}(1-\ln\frac{m^{2}}{\mud^{2}}-r\ln|\frac{1+r}{1-r}|)+2L(\mud)
& \left(q^{2}<0\right)\\ \displaystyle
-\frac{1}{16\pi^{2}}(1-\ln\frac{m^{2}}{\mud^{2}}-2r\,
\arctan\frac{1}{r})+2L(\mud) & (0<q^{2}<4m^{2})\\ \displaystyle
-\frac{1}{16\pi^{2}}(1-\ln\frac{m^{2}}{\mud^{2}}-r\ln|\frac{1+r}{1-r}|+i\pi
r)+2L(\mud) & (q^{2}>4m^{2})
\end{cases},
\end{eqnarray}
where $r=\sqrt{|1-4m^{2}/q^{2}|}$.

\subsection{Integrals with one baryon propagator and one meson
propagator}

\begin{equation}
i\int\frac{d^{d}l
\,\mud^{4-d}}{(2\pi)^{d}}\frac{[1,l_{\alpha},l_{\alpha}l_{\beta}]}{(l^{2}-m^{2}+i\epsilon)(\omega+v\cdot
l+i\epsilon)}
=[J_{0}(\omega),v_{\alpha}J_{1}(\omega),g_{\alpha\beta}J_{2}(\omega)+v_{\alpha}v_{\beta}J_3(\omega)],
\end{equation}
\begin{equation}
J_{0}(\omega)=\begin{cases} \displaystyle
\frac{-\omega}{8\pi^{2}}(1-\ln\frac{m^{2}}{\mud^{2}})+\frac{\sqrt{\omega^{2}-m^{2}}}{4\pi^{2}}({\rm
arccosh}\frac{\omega}{m}-i\pi)+4\omega L(\mud) & (\omega>m)\\
\displaystyle
\frac{-\omega}{8\pi^{2}}(1-\ln\frac{m^{2}}{\mud^{2}})+\frac{\sqrt{m^{2}-\omega^{2}}}{4\pi^{2}}\arccos\frac{-\omega}{m}+4\omega
L(\mud) & (\omega^{2}<m^{2})\\ \displaystyle
\frac{-\omega}{8\pi^{2}}(1-\ln\frac{m^{2}}{\mud^{2}})-\frac{\sqrt{\omega^{2}-m^{2}}}{4\pi^{2}}{\rm
arccosh}\frac{-\omega}{m}+4\omega L(\mud) & (\omega<-m)
\end{cases}
\end{equation}

\begin{eqnarray}
J_{1}(\omega)&=&-\omega J_{0}(\omega)+\Delta\\
J_2(\omega)&=&\frac{1}{d-1}[(m^2-\omega^2)J_0(\omega)+\omega\Delta]\\
J_3(\omega)&=&-\omega J_1(\omega)-J_2(\omega)
\end{eqnarray}

\subsection{Integrals with two baryon propagators and one meson propagator}

\begin{eqnarray}
i\int\frac{d^{d}l
\,\mud^{4-d}}{(2\pi)^{d}}\frac{[1,l_{\alpha},l_{\alpha}l_{\beta}]}{(l^{2}-m^{2}+i\epsilon)(v\cdot
l+i\epsilon)(\omega+v\cdot l+i\epsilon)}
=[\Gamma_{0}(\omega),v_{\alpha}\Gamma_{1}(\omega),g_{\alpha\beta}\Gamma_{2}(\omega)+v_{\alpha}v_{\beta}\Gamma_3(\omega)]
\qquad \omega\neq 0
\end{eqnarray}

\begin{eqnarray}
\Gamma_{i}(\omega)=\frac{1}{\omega}[J_i(0)-J_i(\omega)]
\end{eqnarray}

\begin{eqnarray}
i\int\frac{d^{d}l\,\mud^{4-d}}{(2\pi)^{d}}\frac{[1,l_{\alpha},l_{\alpha}l_{\beta}]}{(l^{2}-m^{2}+i\epsilon)(\omega+v\cdot
l+i\epsilon)^2} =-[\frac{\partial}{\partial
\omega}J_{0}(\omega),v_{\alpha}\frac{\partial}{\partial
\omega}J_{1}(\omega),g_{\alpha\beta}\frac{\partial}{\partial
\omega}J_{2}(\omega)+v_{\alpha}v_{\beta}\frac{\partial}{\partial
\omega}J_3(\omega)]
\end{eqnarray}

\subsection{Integrals with one baryon propagator and two meson propagators}

\begin{eqnarray*}
i\int\frac{d^{d}l\,\mud^{4-d}}{(2\pi)^{d}}\frac{[1,l_{\alpha},l_{\alpha}l_{\beta},l_{\nu}l_{\alpha}l_{\beta}]}{(l^{2}-m^{2}+i\epsilon)((l+q)^{2}-m^{2}+i\epsilon)(\omega+v\cdot
l+i\epsilon)} & = &
[L_{0}(\omega),L_{\alpha},L_{\alpha\beta},L_{\nu\alpha\beta}],\beta=\omega-v\cdot
q
\end{eqnarray*}

\begin{eqnarray}
L_{0}(\omega)& = & \protect\begin{cases} \displaystyle
\frac{1}{8\pi^{2}v\cdot q}\left\{\frac{1}{2}\left[\left({\rm
arccosh}\frac{\beta}{m}\right)^2-\left({\rm
arccosh}\frac{\omega}{m}\right)^2\right]-i\pi \ln
\frac{\sqrt{\beta^{2}-m^{2}}+\beta}{\sqrt{\omega^{2}-m^{2}}+\omega}\right\}
& (\beta>m)\\ \displaystyle \frac{1}{16\pi^{2}v\cdot
q}\left[\left({\rm arccos}\frac{-\omega}{m}\right)^2-\left({\rm
arccos}\frac{-\beta}{m}\right)^2\right] &
(\beta^{2}<m^{2})\protect\\ \displaystyle \frac{1}{16\pi^{2}v\cdot
q}\left[\left({\rm arccosh}\frac{-\beta}{m}\right)^2-\left({\rm
arccosh}\frac{-\omega}{m}\right)^2\right] & (\beta<-m)
\protect\end{cases}.\\
L_{\alpha}& = & n^{\rmI}_{1}q_{\alpha}+n^{\rmI}_{2}v_{\alpha}\\
L_{\alpha\beta} & = &
n^{\rmII}_{1}g_{\alpha\beta}+n^{\rmII}_{2}q_{\alpha}q_{\beta}+n^{\rmII}_{3}v_{\alpha}v_{\beta}
 +n^{\rmII}_{4}v_{\alpha}q_{\beta}+n^{\rmII}_{5}q_{\alpha}v_{\beta}\\
 L_{\nu\alpha\beta} & = & n^{\rmIII}_{1}q_{\nu}q_{\alpha}q_{\beta}+n^{\rmIII}_{2}q_{\nu}q_{\alpha}v_{\beta}
 +n^{\rmIII}_{3}q_{\nu}q_{\beta}v_{\alpha}+n^{\rmIII}_{4}q_{\alpha}q_{\beta}v_{\nu}
 +n^{\rmIII}_{5}q_{\nu}g_{\alpha\beta}\nonumber\\
 &  & +n^{\rmIII}_{6}q_{\beta}g_{\nu\alpha}+n^{\rmIII}_{7}q_{\alpha}g_{\nu\beta}+n^{\rmIII}_{8}q_{\nu}v_{\alpha}v_{\beta}
 +n^{\rmIII}_{9}q_{\alpha}v_{\nu}v_{\beta}
 +n^{\rmIII}_{10}q_{\beta}v_{\nu}v_{\alpha}\nonumber\\
 &  & +n^{\rmIII}_{11}g_{\nu\beta}v_{\alpha}+n^{\rmIII}_{12}g_{\nu\alpha}v_{\beta}
 +n^{\rmIII}_{13}g_{\alpha\beta}v_{\nu}+n^{\rmIII}_{14}v_{\nu}v_{\alpha}v_{\beta}
 \end{eqnarray}

\subsection{Explicit expressions of the scalar functions}

\begin{eqnarray*}
n^{\rmI}_{1} & = & \frac{2(v\cdot q)I_{0}+2(v\cdot q)L_{0}\omega-J_{0}(\beta)+J_{0}(\omega)}{2(v\cdot q)^{2}}\\
n^{\rmI}_{2} & = & \frac{J_{0}(\beta)-J_{0}(\omega)}{2(v\cdot q)}\\
n^{\rmII}_{1} & = & \frac{L_{0}m^{2}2(v\cdot q)+[2(v\cdot q)-\beta+2\omega]J_{0}(\beta)-\omega J_{0}(\omega)}{(2(d-2))(v\cdot q)}\\
n^{\rmII}_{2} & = & \frac{1}{2(d-2)(v\cdot q)^{3}}[(v\cdot q)\left(2\left(L_{0}\left((d-2)\omega^{2}+m^{2}\right)-I_{0}(d-2)\omega\right)-I_{0}(d-2)(v\cdot q)\right)\\
 &  &+J_{0}(\beta)(2(v\cdot q)-(d-1)(\beta-2\omega))+(\omega-d\omega)J_{0}(\omega)]\\
n^{\rmII}_{3} & = & \frac{\omega J_{0}(\omega)-\beta J_{0}(\beta)}{2(v\cdot q)}\\
n^{\rmII}_{4,5} & = & \frac{-J_{0}(\beta)(\beta-\beta d+d\omega+2(v\cdot q))-2(v\cdot q)L_{0}m^{2}+\omega J_{0}(\omega)}{2(d-2)(v\cdot q)^{2}}\\
n^{\rmIII}_{1} & = & \frac{1}{2(d-2)(d-1)(v\cdot q)^{4}}[-\beta d^{2}\Delta+d^{2}\Delta\omega+2I_{3}d^{2}(v\cdot q)^{3}+I_{0}d^{2}(v\cdot q)^{2}\omega-2d^{2}(v\cdot q)L_{0}\omega^{3}\\
 &  & +J_{0}(\beta)\left(-3\beta\omega+\beta^{2}d^{2}+3\beta d^{2}\omega-3d^{2}\omega^{2}+\beta^{2}d-6(d-1)(v\cdot q)^{2}+(2-4d)m^{2}+3d\omega^{2}\right)\\
 &  &+2I_{0}d^{2}(v\cdot q)\omega^{2}+J_{0}(\omega)\left((4d-2)m^{2}-\left(d^{2}+4d-3\right)\omega^{2}\right)-\beta d\Delta+d\Delta\omega-6I_{3}d(v\cdot q)^{3}\\
 &  &-3I_{0}d(v\cdot q)^{2}\omega-2d\Delta (v\cdot q)+6d(v\cdot q)L_{0}m^{2}\omega+6d(v\cdot q)L_{0}\omega^{3}+2I_{0}d(v\cdot q)m^{2}-6I_{0}d(v\cdot q)\omega^{2}\\
 &  & +4I_{3}(v\cdot q)^{3}+2I_{0}(v\cdot q)^{2}\omega-2\Delta (v\cdot q)-6(v\cdot q)L_{0}m^{2}\omega-4(v\cdot q)L_{0}\omega^{3}-10I_{0}(v\cdot q)m^{2}+4I_{0}(v\cdot q)\omega^{2}]\\
n^{\rmIII}_{2,3,4} & = & \frac{1}{2(d-2)(d-1)(v\cdot q)^{3}}[J_{0}(\beta)[4\beta\omega-\beta^{2}d^{2}-2\beta d^{2}\omega+d^{2}\omega^{2}-\beta^{2}d-2\beta d\omega\\
 &  & +4(d-1)(v\cdot q)^{2}-2\beta(d-1)(v\cdot q)+(4d-2)m^{2}+d\omega^{2}-2\omega^{2}]+\beta d^{2}\Delta-d^{2}\Delta\omega\\
 &  &+2J_{0}(\omega)\left(\left(d^{2}+d-1\right)\omega^{2}+(1-2d)m^{2}\right)+\beta d\Delta-d\Delta\omega+3d\Delta (v\cdot q)\\
 &  &-4d(v\cdot q)L_{0}m^{2}\omega-2I_{0}d(v\cdot q)m^{2}+4(v\cdot q)L_{0}m^{2}\omega+8I_{0}(v\cdot q)m^{2}]\\
n^{\rmIII}_{5,6,7} & = & \frac{1}{2(d-2)(d-1)(v\cdot q)^{2}}[J_{0}(\beta)\left(\omega(2\omega-3\beta)-2(d-1)(v\cdot q)^{2}+d\left(\beta^{2}+3\beta\omega-m^{2}-2\omega^{2}\right)\right)\\
&  &-\beta d\Delta+d\Delta\omega+2d(v\cdot q)L_{0}m^{2}\omega+2I_{0}d(v\cdot q)m^{2}+J_{0}(\omega)\left(d\left(m^{2}-2\omega^{2}\right)+\omega^{2}\right)\\
&  &-2\Delta (v\cdot q)-2(v\cdot q)L_{0}m^{2}\omega-6I_{0}(v\cdot q)m^{2}]\\
n^{\rmIII}_{8,9,10}& = & \frac{1}{2(d-2)(d-1)(v\cdot q)^{2}}[J_{0}(\beta)\left(\beta\left(d^{2}(\beta+\omega)+4(d-1)(v\cdot q)+d\omega-2\omega\right)+(2-3d)m^{2}\right)\\
&  &\Delta\left(d^{2}(\omega-\beta)-4(d-1)(v\cdot q)\right)+J_{0}(\omega)\left((3d-2)m^{2}-\left(2d^{2}+d-2\right)\omega^{2}\right)]\\
n^{\rmIII}_{11,12,13} & = & \frac{1}{2(d-2)(d-1)(v\cdot q)}[J_{0}(\beta)\left(d\left(m^{2}-\beta(\beta+2(v\cdot q)+2\omega)\right)+2\beta((v\cdot q)+\omega)\right)\\
&  &\Delta(d(\beta-\omega)+2(d-1)(v\cdot q))-J_{0}(\omega)\left(d\left(m^{2}-3\omega^{2}\right)+2\omega^{2}\right)]\\
n^{\rmIII}_{14} & = & \frac{1}{2(d-1)(v\cdot q)}[d\Delta(\beta-\omega)+J_{0}(\beta)\left(m^{2}-\beta^{2}d\right)+J_{0}(\omega)\left(d\omega^{2}-m^{2}\right)]\\
\end{eqnarray*}

\section{COEFFICIENTS OF THE LOOP CORRECTIONS} \label{appendix-B}

In this appendix, we collect the explicit formulae for the chiral
expansion of the decuplet baryon magnetic moments at
$\mathcal{O}(p^{2})$ in Table \ref{table:beta}  and
$\mathcal{O}(p^{3})$ in Tables \ref{table:cd}, \ref{table:ef},
\ref{table:gh}, \ref{table:mn} and \ref{table:op}, respectively.

\begin{table}
  \centering
\begin{tabular}{c|c|c|c|c}
\toprule[1pt]\toprule[1pt] Process & $\beta_{\mathcal{T}}^{\pi}$ &
$\beta_{\mathcal{T}}^{K}$ & $\beta_{\mathcal{N}}^{\pi}$ &
$\beta_{\mathcal{N}}^{K}$\tabularnewline \midrule[1pt]
$\Delta^{+}\rightarrow p\gamma$ & $\frac{10\sqrt{3}}{9}$ &
$\frac{2\sqrt{3}}{9}$ & $-\frac{2\sqrt{3}}{3}(D+F)$ &
$-\frac{2\sqrt{3}}{3}(D-F)$\tabularnewline

$\Delta^{0}\rightarrow n\gamma$ & $\frac{10\sqrt{3}}{9}$ &
$\frac{2\sqrt{3}}{9}$ & $-\frac{2\sqrt{3}}{3}(D+F)$ &
$-\frac{2\sqrt{3}}{3}(D-F)$\tabularnewline

$\Sigma^{*+}\rightarrow\Sigma^{+}\gamma$ &
$-\frac{2\sqrt{3}}{9}$ & $-\frac{10\sqrt{3}}{9}$ &
$\frac{2\sqrt{3}}{3}(D-F)$ &
$\frac{2\sqrt{3}}{3}(D+F)$\tabularnewline

$\Sigma^{*0}\rightarrow\Sigma^{0}\gamma$ & $0$ &
$\frac{2\sqrt{3}}{3}$ & $0$ & $-\frac{2\sqrt{3}}{3}D$\tabularnewline

$\Sigma^{*0}\rightarrow\Lambda\gamma$ & $-\frac{4}{3}$ &
$-\frac{2}{3}$ & $\frac{4D}{3}$ & $\frac{2}{3}D$\tabularnewline

$\Sigma^{*-}\rightarrow\Sigma^{-}\gamma$ &
$-\frac{2\sqrt{3}}{9}$ & $\frac{2\sqrt{3}}{9}$ &
$\frac{2\sqrt{3}}{3}(D-F)$ &
$-\frac{2\sqrt{3}}{3}(D-F)$\tabularnewline

$\Xi^{*0}\rightarrow\Xi^{0}\gamma$ & $-\frac{2\sqrt{3}}{9}$ &
$-\frac{10\sqrt{3}}{9}$ & $\frac{2\sqrt{3}}{3}(D-F)$ &
$\frac{2\sqrt{3}}{3}(D+F)$\tabularnewline

$\Xi^{*-}\rightarrow\Xi^{-}\gamma$ & $-\frac{2\sqrt{3}}{9}$ &
$\frac{2\sqrt{3}}{9}$ & $\frac{2\sqrt{3}}{3}(D-F)$ &
$-\frac{2\sqrt{3}}{3}(D-F)$\tabularnewline
\bottomrule[1pt]\bottomrule[1pt]
\end{tabular}
\caption{The coefficients of the loop corrections to the
decuplet to octet baryon transition magnetic moments from Figs.
\ref{fig:allloop}(a) and \ref{fig:allloop}(b). The subscripts
``$\mathcal T$'' and ``$\mathcal N$'' denote the decuplet and octet
baryon within the loop while the superscripts denote the
pseudoscalar meson. What calls for special attention is that the pion loop coefficients of $\Sigma^{*0}\rightarrow\Sigma^{0}\gamma$ channel are zero. The reason is that the pion loop contributions from different intermediate states cancel each
other. In other words, the pion loop contributions with the
intermediate baryons $\Sigma^{*+}$ and $\Sigma^{*-}$, $\Sigma^{+}$ and $\Sigma^{-}$ cancel each
other due to the exact SU(2) flavor symmetry.}\label{table:beta}
\end{table}

\begin{table}
  \centering
\begin{tabular}{c|c|c|c|c|c|c}
\toprule[1pt]\toprule[1pt] Process & $\gamma_{c}^{\pi}$ &
$\gamma_{c}^{K}$ & $\gamma_{c}^{\eta}$ & $\gamma_{d}^{\pi}$ &
$\gamma_{d}^{K}$ & $\gamma_{d}^{\eta}$\tabularnewline \midrule[1pt]
$\Delta^{+}\rightarrow p\gamma$ & $\frac{2}{\sqrt{3}}$ &
$\frac{1}{\sqrt{3}}$ & 0 &
$-\frac{4\sqrt{3}}{3}(\tilde{g}_{t1}+\tilde{g}_{t2})$ &
$-\frac{2\sqrt{3}}{3}\tilde{g}_{t1}$ & 0\tabularnewline

$\Delta^{0}\rightarrow n\gamma$ & $\frac{2}{\sqrt{3}}$ &
$\frac{1}{\sqrt{3}}$ & 0 &
$-\frac{4\sqrt{3}}{3}(\tilde{g}_{t1}+\tilde{g}_{t2})$ &
$-\frac{2\sqrt{3}}{3}\tilde{g}_{t1}$ & 0\tabularnewline

$\Sigma^{*+}\rightarrow\Sigma^{+}\gamma$ & $-\frac{1}{\sqrt{3}}$
& $-\frac{2}{\sqrt{3}}$ & 0 & $\frac{2\sqrt{3}}{3}\tilde{g}_{t1}$ &
$\frac{4\sqrt{3}}{3}\tilde{g}_{t1}+\frac{4\sqrt{3}}{3}\tilde{g}_{t2}$
& 0\tabularnewline

$\Sigma^{*0}\rightarrow\Sigma^{0}\gamma$ & 0 &
$\frac{\sqrt{3}}{2}$ & 0 & $0$ &
$-\sqrt{3}\tilde{g}_{t1}-\frac{2\sqrt{3}}{3}\tilde{g}_{t2}$ &
0\tabularnewline

$\Sigma^{*0}\rightarrow\Lambda\gamma$ & -1 & $-\frac{1}{2}$ & 0
& $2\tilde{g}_{t1}+\frac{4}{3}\tilde{g}_{t2}$ &
$\tilde{g}_{t1}+\frac{4}{3}\tilde{g}_{t2}$ & 0\tabularnewline

$\Sigma^{*-}\rightarrow\Sigma^{-}\gamma$ & $-\frac{1}{\sqrt{3}}$
& $\frac{1}{\sqrt{3}}$ & 0 & $\frac{2\sqrt{3}}{3}\tilde{g}_{t1}$ &
$-\frac{2\sqrt{3}}{3}\tilde{g}_{t1}$ & 0\tabularnewline

$\Xi^{*0}\rightarrow\Xi^{0}\gamma$ & $-\frac{1}{\sqrt{3}}$ &
$-\frac{2}{\sqrt{3}}$ & 0 & $\frac{2\sqrt{3}}{3}\tilde{g}_{t1}$ &
$\frac{4\sqrt{3}}{3}\tilde{g}_{t1}+\frac{4\sqrt{3}}{3}\tilde{g}_{t2}$
& 0\tabularnewline

$\Xi^{*-}\rightarrow\Xi^{-}\gamma$ & $-\frac{1}{\sqrt{3}}$ &
$\frac{1}{\sqrt{3}}$ & 0 & $\frac{2\sqrt{3}}{3}\tilde{g}_{t1}$ &
$-\frac{2\sqrt{3}}{3}\tilde{g}_{t1}$ & 0\tabularnewline
\bottomrule[1pt]\bottomrule[1pt]
\end{tabular}
\caption{The coefficients of the loop corrections to the
decuplet to octet baryon transition magnetic moments from Figs.
\ref{fig:allloop}(c) and \ref{fig:allloop}(d).} \label{table:cd}
\end{table}

\begin{table}
  \centering
\begin{tabular}{c|c|c|c|c|c|c}
\toprule[1pt]\toprule[1pt] Process & $\gamma_{a\mathcal{T}}^{\pi}$ &
$\gamma_{a\mathcal{T}}^{K}$ & $\gamma_{a\mathcal{T}}^{\eta}$ &
$\gamma_{a\mathcal{N}}^{\pi}$ & $\gamma_{a\mathcal{N}}^{K}$ &
$\gamma_{a\mathcal{N}}^{\eta}$\tabularnewline \midrule[1pt]
$\Delta^{+}\rightarrow p\gamma$ & $\frac{10\sqrt{3}}{9}b$ &
$\frac{2\sqrt{3}}{9}b$ & 0 & $-\frac{2}{\sqrt{3}}(b_{D}+b_{F})(D+F)$
& $-\frac{2}{3\sqrt{3}}[b_{D}(D+3F)+3b_{F}(D-F)]$ & 0\tabularnewline

$\Delta^{0}\rightarrow n\gamma$ & $\frac{10\sqrt{3}}{9}b$ &
$\frac{2\sqrt{3}}{9}b$ & $0$ &
$-\frac{2}{\sqrt{3}}(b_{D}+b_{F})(D+F)$ &
$-\frac{2}{3\sqrt{3}}[b_{D}(D+3F)+3b_{F}(D-F)]$ & 0\tabularnewline

$\Sigma^{*+}\rightarrow\Sigma^{+}\gamma$ &
$\frac{2\sqrt{3}}{9}b$ & $\frac{-14\sqrt{3}}{9}b$ & 0 &
$\frac{2}{3\sqrt{3}}(2b_{D}+3b_{F})F$ &
$\frac{2}{3\sqrt{3}}[b_{D}(3D+F)+3b_{F}(D-F)]$ &
$\frac{2}{3\sqrt{3}}(b_{D}+3b_{F})D$\tabularnewline

$\Sigma^{*0}\rightarrow\Sigma^{0}\gamma$ & 0 &
$\frac{2\sqrt{3}}{3}b$ & 0 & $-\frac{2}{3\sqrt{3}}b_{D}(D+2F)$ &
$-\frac{2}{3\sqrt{3}}(b_{D}F+3b_{F}D)$ &
$-\frac{2}{3\sqrt{3}}b_{D}D$\tabularnewline

$\Sigma^{*0}\rightarrow\Lambda\gamma$ & $-\frac{4}{3}b$ &
$-\frac{2}{3}b$ &0  & $\frac{4}{3}b_{F}D$ &
$2b_{D}F+\frac{2}{3}b_{F}D$ & $\frac{2}{3}b_{D}D$\tabularnewline

$\Sigma^{*-}\rightarrow\Sigma^{-}\gamma$ &
$\frac{2\sqrt{3}}{9}b$ & $-\frac{2\sqrt{3}}{9}b$ &  0&
$\frac{1}{3\sqrt{3}}[6b_{F}F-4b_{D}(D+F)]$ &
$\frac{1}{3\sqrt{3}}[b_{D}(6D-2F)-6b_{F}(D+F)]$ &
$\frac{2}{\sqrt{3}}(-\frac{1}{3}b_{D}+b_{F})D$\tabularnewline

$\Xi^{*0}\rightarrow\Xi^{0}\gamma$ & $-\frac{\sqrt{6}}{9}b$ &
$-\frac{10\sqrt{3}}{9}b$ & $0$ & $\frac{2}{\sqrt{3}}b_{F}(D-F)$ &
$\frac{1}{\sqrt{3}}(b_{D}+2b_{F})(D+F)$ &
$\frac{2}{3\sqrt{3}}b_{D}(D+3F)$\tabularnewline

$\Xi^{*-}\rightarrow\Xi^{-}\gamma$ & $\frac{\sqrt{3}}{9}b$ &
$\frac{2\sqrt{3}}{9}b$ & $-\frac{\sqrt{3}}{3}b$ &
$-\frac{1}{\sqrt{3}}(b_{D}+b_{F})(D-F)$ &
$\frac{1}{3\sqrt{3}}[-b_{D}(D+9F)+6b_{F}(D+F)]$ &
$\frac{1}{3\sqrt{3}}(b_{D}-3b_{F})(D+3F)$\tabularnewline
\bottomrule[1pt]\bottomrule[1pt]
\end{tabular}
\caption{The coefficients of the loop corrections to the
decuplet to octet baryon transition magnetic moments from Figs.
\ref{fig:allloop}(e) and \ref{fig:allloop}(f).} \label{table:ef}
\end{table}

\begin{table}
  \centering
\begin{tabular}{c|c|c|c|c|c|c}
\toprule[1pt]\toprule[1pt] Process &
$\gamma_{a\mathcal{T}\mathcal{N}}^{\pi}$ &
$\gamma_{a\mathcal{T}\mathcal{N}}^{K}$ &
$\gamma_{a\mathcal{T}\mathcal{N}}^{\eta}$ &
$\gamma_{a\mathcal{N}\mathcal{T}}^{\pi}$ &
$\gamma_{a\mathcal{N}\mathcal{T}}^{K}$ &
$\gamma_{a\mathcal{N}\mathcal{T}}^{\eta}$\tabularnewline
\midrule[1pt] $\Delta^{+}\rightarrow p\gamma$ &
$\frac{10}{3\sqrt{3}}(D+F)$ & $\frac{8}{3\sqrt{3}}F$ &
$-\frac{2}{3\sqrt{3}}(D-3F)$ & $\frac{4}{3\sqrt{3}}$ &
$\frac{2}{3\sqrt{3}}$ & 0\tabularnewline

$\Delta^{0}\rightarrow n\gamma$ & $\frac{10}{3\sqrt{3}}(D+F)$ &
$\frac{8}{3\sqrt{3}}F$ & $-\frac{2}{3\sqrt{3}}(D-3F)$ &
$\frac{4}{3\sqrt{3}}$ & $\frac{2}{3\sqrt{3}}$ & 0\tabularnewline

$\Sigma^{*+}\rightarrow\Sigma^{+}\gamma$ &
$-\frac{4}{3\sqrt{3}}(D+3F)$ & $-\frac{4}{3\sqrt{3}}(D+3F)$ & 0 & 0
& 0 & $-\frac{2}{\sqrt{3}}$\tabularnewline

$\Sigma^{*0}\rightarrow\Sigma^{0}\gamma$ &
$\frac{8}{3\sqrt{3}}F$ & $\frac{4}{3\sqrt{3}}(D+F)$ & 0 &
$\frac{2}{3\sqrt{3}}$ & $-\frac{2}{3\sqrt{3}}$ &
$\frac{1}{\sqrt{3}}$\tabularnewline

$\Sigma^{*0}\rightarrow\Lambda\gamma$ & $-\frac{8}{9}D$ &
$-\frac{4}{9}(D+9F)$ & 0 & $-\frac{1}{3}$ & $-\frac{2}{3}$ &
0\tabularnewline

$\Sigma^{*-}\rightarrow\Sigma^{-}\gamma$ &
$-\frac{4}{3\sqrt{3}}(D-F)$ & $\frac{4}{3\sqrt{3}}(D-F)$ & 0 &
$\frac{4}{3\sqrt{3}}$ & $-\frac{4}{3\sqrt{3}}$ & 0\tabularnewline

$\Xi^{*0}\rightarrow\Xi^{0}\gamma$ & $\frac{2}{3\sqrt{3}}(D-F)$
& $-\frac{8}{3\sqrt{3}}(D+2F)$ & $\frac{2}{3\sqrt{3}}(D+3F)$ &
$-\frac{2}{3\sqrt{3}}$ & $\frac{2}{3\sqrt{3}}$ &
$-\frac{2}{\sqrt{3}}$\tabularnewline

$\Xi^{*-}\rightarrow\Xi^{-}\gamma$ & $-\frac{4}{3\sqrt{3}}(D-F)$
& $\frac{4}{3\sqrt{3}}(D-F)$ & 0 & $\frac{4}{3\sqrt{3}}$ &
$-\frac{4}{3\sqrt{3}}$ & 0\tabularnewline
\bottomrule[1pt]\bottomrule[1pt]
\end{tabular}
\caption{The coefficients of the loop corrections to the
decuplet to octet baryon transition magnetic moments from Figs.
\ref{fig:allloop}(g) and \ref{fig:allloop}(h).} \label{table:gh}
\end{table}

\begin{table}
  \centering
\begin{tabular}{c|c|c|c|c|c|c}
\toprule[1pt]\toprule[1pt] Baryons & $\gamma_{f\mathcal{T}10}^{\pi}$
& $\gamma_{f\mathcal{T}10}^{K}$ & $\gamma_{f\mathcal{T}10}^{\eta}$ &
$\gamma_{f\mathcal{T}8}^{\pi}$ & $\gamma_{f\mathcal{T}8}^{K}$ &
$\gamma_{f\mathcal{T}8}^{\eta}$\tabularnewline \midrule[1pt]
$\Delta^{+}$ & $\frac{5}{3}$ & $\frac{2}{3}$ & $\frac{1}{3}$ & $2$ &
$2$ & 0\tabularnewline

$\Delta^{0}$ & $\frac{5}{3}$ & $\frac{2}{3}$ & $\frac{1}{3}$ & $2$ &
$2$ & 0\tabularnewline

$\Delta^{-}$ & $\frac{5}{3}$ & $\frac{2}{3}$ & $\frac{1}{3}$ & $2$ &
$2$ & 0\tabularnewline

$\Sigma^{*+}$ & $\frac{8}{9}$ & $\frac{16}{9}$ & 0 & $\frac{5}{3}$ &
$\frac{4}{3}$ & 1\tabularnewline

$\Sigma^{*0}$ & $\frac{8}{9}$ & $\frac{16}{9}$ & 0 & $\frac{5}{3}$ &
$\frac{4}{3}$ & 1\tabularnewline

$\Sigma^{*-}$ & $\frac{8}{9}$ & $\frac{16}{9}$ & 0 & $\frac{5}{3}$ &
$\frac{4}{3}$ & 1\tabularnewline

$\Xi^{*0}$ & $\frac{1}{3}$ & $2$ & $\frac{1}{3}$ & $1$ & $2$ &
1\tabularnewline

$\Xi^{*-}$ & $\frac{1}{3}$ & $2$ & $\frac{1}{3}$ & $1$ & $2$ &
1\tabularnewline \bottomrule[1pt]\bottomrule[1pt]
\end{tabular}
\caption{The coefficients of the loop corrections to the
decuplet to octet baryon transition magnetic moments from Figs.
\ref{fig:allloop}(m) and \ref{fig:allloop}(n).} \label{table:mn}
\end{table}

\begin{table}
  \centering
\begin{tabular}{c|c|c|c|c|c|c}
\toprule[1pt]\toprule[1pt] Baryons & $\gamma_{f\mathcal{N}8}^{\pi}$
& $\gamma_{f\mathcal{N}8}^{K}$ & $\gamma_{f\mathcal{N}8}^{\eta}$ &
$\gamma_{f\mathcal{N}10}^{\pi}$ & $\gamma_{f\mathcal{N}10}^{K}$ &
$\gamma_{f\mathcal{N}10}^{\eta}$\tabularnewline \midrule[1pt] $p$ &
$3(D+F)^{2}$ & $\frac{10}{3}D^{2}-4DF+6F^{2}$ &
$\frac{1}{3}(D-3F)^{2}$ & $4$ & $1$ & 0\tabularnewline

$n$ & $3(D+F)^{2}$ & $\frac{10}{3}D^{2}-4DF+6F^{2}$ &
$\frac{1}{3}(D-3F)^{2}$ & $4$ & $ 1$ & 0\tabularnewline

$\Sigma^{+}$ & $\frac{4}{3}D^{2}+8F^{2}$ & $4(D^{2}+F^{2})$ &
$\frac{4}{3}D^{2}$ & $\frac{2}{3}$ & $\frac{10}{3}$ &
1\tabularnewline

$\Sigma^{0}$ & $\frac{4}{3}D^{2}+8F^{2}$ & $4(D^{2}+F^{2})$ &
$\frac{4}{3}D^{2}$ & $\frac{2}{3}$ & $\frac{10}{3}$ &
1\tabularnewline

$\Lambda$ & $4D^{2}$ & $\frac{4}{3}(D^{2}+9F^{2})$ &
$\frac{4}{3}D^{2}$ & $3$ & $2$ & 0\tabularnewline

$\Sigma^{-}$ & $\frac{4}{3}D^{2}+8F^{2}$ & $4(D^{2}+F^{2})$ &
$\frac{4}{3}D^{2}$ & $\frac{2}{3}$ & $\frac{10}{3}$ &
1\tabularnewline

$\Xi^{0}$ & $3(D-F)^{2}$ & $\frac{10}{3}D^{2}+4DF+6F^{2}$ &
$\frac{1}{3}(D+3F)^{2}$ & $1$ & $3$ & 1\tabularnewline

$\Xi^{-}$ & $3(D-F)^{2}$ & $\frac{10}{3}D^{2}+4DF+6F^{2}$ &
$\frac{1}{3}(D+3F)^{2}$ & $1$ & $3$ & 1\tabularnewline
\bottomrule[1pt]\bottomrule[1pt]
\end{tabular}
\caption{The coefficients of the loop corrections to the
decuplet to octet baryon transition magnetic moments from Figs.
\ref{fig:allloop}(o) and \ref{fig:allloop}(p).} \label{table:op}
\end{table}

\section{Transition amplitudes and decay width} \label{appendix-C}

We collect the M1 and E2 amplitudes and decay width of the decuplet
to octet baryon transitions in this appendix.
\begin{table}
\centering
\begin{tabular}{c|ccc}
\toprule[1pt]\toprule[1pt] Process(fit A) &
$f_{M1}$/$\rm{GeV^{-\frac{1}{2}}}$&
$f_{E2}$/$\rm{GeV^{-\frac{1}{2}}}$& Decay width $\Gamma$/\rm{MeV}
\tabularnewline \midrule[1pt] $\Delta\rightarrow N\gamma$ &
$-0.317+0.024i$ & $0.008+0.018i$ & 0.73\tabularnewline

$\Sigma^{*+}\rightarrow\Sigma^{+}\gamma$ & $0.246-0.022i$ & $-0.003-
0.002i $ & 0.25\tabularnewline

$\Sigma^{*0}\rightarrow\Sigma^{0}\gamma$ & $-0.129+0.010i$ &$0.001$
& 0.07\tabularnewline

$\Sigma^{*0}\rightarrow\Lambda\gamma$ & $0.255+0.012i$
&$-0.005-0.009i$ &0.43\tabularnewline

$\Sigma^{*-}\rightarrow \Sigma^{-}\gamma$ & $-0.012-0.002i$ & $
-0.001- 0.002i$ &$5.79\times10^{-4}$\tabularnewline

$\Xi^{*0}\rightarrow\Xi^{0}\gamma$ &$0.282+0.023i$ & $-0.003- 0.002i
$ & 0.41 \tabularnewline

$\Xi^{*-}\rightarrow \Xi^{-}\gamma$ & $0.010+0.002i$ &$-0.001 -
0.002i $& $5.25\times10^{-4}$\tabularnewline
 \bottomrule[1pt]\bottomrule[1pt]
\end{tabular}
 \caption{M1 and E2 amplitudes and decay width of the decuplet to octet baryons in fit A.}
\label{table:amplitudes and decay widthA}
\end{table}

\begin{table}
\centering
\begin{tabular}{c|cccc}
\toprule[1pt]\toprule[1pt] Process(fit B) &
$f_{M1}$/$\rm{GeV^{-\frac{1}{2}}}$&
$f_{E2}$/$\rm{GeV^{-\frac{1}{2}}}$& Decay width $\Gamma$/\rm{MeV} &
$\Gamma_{\rm exp}$/\rm{MeV} \tabularnewline \midrule[1pt]
$\Delta\rightarrow N\gamma$ & $-0.321+0.024i$ & $0.008+0.018i$ &
0.75&0.70\tabularnewline

$\Sigma^{*+}\rightarrow\Sigma^{+}\gamma$ & $0.244-0.022i$ & $-0.003-
0.002i $ & 0.25&0.25\tabularnewline

$\Sigma^{*0}\rightarrow\Sigma^{0}\gamma$ & $-0.128+0.010i$ &$0.001$
& 0.07&\textemdash{}\tabularnewline

$\Sigma^{*0}\rightarrow\Lambda\gamma$ & $0.249+0.012i$
&$-0.005-0.009i$ &0.42 & 0.45\tabularnewline

$\Sigma^{*-}\rightarrow \Sigma^{-}\gamma$ & $-0.012-0.002i$ & $
-0.001- 0.002i$ &$6.39\times10^{-4}$ &  $<9.36\times10^{-3}$
\tabularnewline

$\Xi^{*0}\rightarrow\Xi^{0}\gamma$ &$0.236+0.023i$ & $-0.003- 0.002i
$ & 0.29 & $<0.38$\tabularnewline

$\Xi^{*-}\rightarrow \Xi^{-}\gamma$ & $0.010+0.002i$ &$-0.001 -
0.002i $& $5.25\times10^{-4}$ & $<0.38$ \tabularnewline
 \bottomrule[1pt]\bottomrule[1pt]
\end{tabular}
 \caption{M1 and E2 amplitudes and decay width of the decuplet to octet baryons in fit B. All the experimental values are from PDG~\cite{Patrignani}.}
\label{table:amplitudes and decay widthB}
\end{table}

\section{Transition magnetic moments in quark model} \label{appendix-D}

We collect the decuplet
to octet baryon transition magnetic moments in quark model in this appendix. The transition moments are obtained by sandwiching Eq.~(\ref{quarkmodel}) between the decuplet and octet
baryon states.
\begin{eqnarray}
\overrightarrow{\mu}  =\sum_{i}\mu_{i}\overrightarrow{\sigma^{i}}, \label{quarkmodel}
\end{eqnarray}
where
\begin{eqnarray}
\mu_{i}=\frac{e_i}{2m_i},  i=u,\hspace{0.1em} d, \hspace{0.1em} s.
\end{eqnarray}
The expressions for the decuplet
to octet baryon transition magnetic moments are given in Table~\ref{table:quarkmodel}.
\begin{table}
\centering
\begin{tabular}{c|c}
\toprule[1pt]\toprule[1pt] Process&
Expressions \tabularnewline \midrule[1pt]
$\Delta\rightarrow N\gamma$ & $\frac{2\sqrt{2}}{3}(\mu_{d}-\mu_{u})$\tabularnewline

$\Sigma^{*+}\rightarrow\Sigma^{+}\gamma$ & $\frac{2\sqrt{2}}{3}(\mu_{u}-\mu_{s})$ \tabularnewline

$\Sigma^{*0}\rightarrow\Sigma^{0}\gamma$ & $\frac{\sqrt{2}}{3}(-\mu_{u}-\mu_{d}+2\mu{s})$\tabularnewline

$\Sigma^{*0}\rightarrow\Lambda\gamma$& $\frac{\sqrt{6}}{3}(\mu_{u}-\mu_{d})$\tabularnewline

$\Sigma^{*-}\rightarrow \Sigma^{-}\gamma$ &$\frac{2\sqrt{2}}{3}(\mu_{s}-\mu_{d})$
\tabularnewline

$\Xi^{*0}\rightarrow\Xi^{0}\gamma$ &$\frac{2\sqrt{2}}{3}(\mu_{u}-\mu_{s})$\tabularnewline

$\Xi^{*-}\rightarrow \Xi^{-}\gamma$ & $\frac{2\sqrt{2}}{3}(\mu_{d}-\mu_{s})$\tabularnewline
 \bottomrule[1pt]\bottomrule[1pt]
\end{tabular}
 \caption{Decuplet
to octet baryon transition magnetic moments in quark model.}
\label{table:quarkmodel}
\end{table}

\end{appendix}

\vfil \thispagestyle{empty}


\newpage
\begin{thebibliography}{99}

\bibitem{Adkins:1983ya}
  G.~S.~Adkins, C.~R.~Nappi and E.~Witten,
  Nucl.\ Phys.\ B {\bf 228}, 552 (1983).

\bibitem{Cohen:1986va}
  T.~D.~Cohen and W.~Broniowski,
  Phys.\ Rev.\ D {\bf 34}, 3472 (1986).

\bibitem{Leinweber:1990dv}
  D.~B.~Leinweber, R.~M.~Woloshyn and T.~Draper,
  Phys.\ Rev.\ D {\bf 43}, 1659 (1991).


  \bibitem{Leinweber:1992hy}
  D.~B.~Leinweber, T.~Draper and R.~M.~Woloshyn,
  Phys.\ Rev.\ D {\bf 46}, 3067 (1992).


\bibitem{Jenkins:1992pi}
  E.~E.~Jenkins, M.~E.~Luke, A.~V.~Manohar and M.~J.~Savage,
  Phys.\ Lett.\ B {\bf 302}, 482 (1993), [Erratum: Phys.\ Lett.\ B {\bf 388}, 866 (1996)].  



\bibitem{Butler:1993ej}
  M.~N.~Butler, M.~J.~Savage and R.~P.~Springer,
  Phys.\ Rev.\ D {\bf 49}, 3459 (1994)  


\bibitem{Banerjee:1995wz}
  M.~K.~Banerjee and J.~Milana,
  Phys.\ Rev.\ D {\bf 54}, 5804 (1996).


\bibitem{Meissner:1997hn}
  U.~G.~Meissner and S.~Steininger,
  Nucl.\ Phys.\ B {\bf 499}, 349 (1997).

\bibitem{Zhu:1998aj}
  S.~L.~Zhu, W.~Y.~P.~Hwang and Z.~s.~Yang,
  Phys.\ Rev.\ D {\bf 57}, 1527 (1998)
.


\bibitem{Meissner:1999hk}
  U.~G.~Meissner,
  Nucl.\ Phys.\ A {\bf 666}, 51 (2000)
  .

\bibitem{Zhu:2000gn}
  S.~L.~Zhu, S.~J.~Puglia, B.~R.~Holstein and M.~J.~Ramsey-Musolf,
  Phys.\ Rev.\ D {\bf 62}, 033008 (2000)
  .




\bibitem{puglia2}
S.~J.~Puglia, and M.~J.~Ramsey-Musolf, Phys. Rev. D62 (2000) 034010.

\bibitem{Kubis:2000aa}
  B.~Kubis and U.~G.~Meissner,
  Eur.\ Phys.\ J.\ C {\bf 18}, 747 (2001)
  .


\bibitem{Puglia:2000jy}
  S.~J.~Puglia, M.~J.~Ramsey-Musolf and S.~L.~Zhu,
  Phys.\ Rev.\ D {\bf 63}, 034014 (2001).  




\bibitem{Savage:2001dy}
  M.~J.~Savage,
  Nucl.\ Phys.\ A {\bf 700}, 359 (2002).




\bibitem{Arndt:2003we}
  D.~Arndt and B.~C.~Tiburzi,
  Phys.\ Rev.\ D {\bf 68}, 114503 (2003)
  [Erratum: Phys.\ Rev.\ D {\bf 69}, 059904 (2004)].




\bibitem{Cloet:2003jm}
  I.~C.~Cloet, D.~B.~Leinweber and A.~W.~Thomas,
  Phys.\ Lett.\ B {\bf 563}, 157 (2003).

\bibitem{Gockeler:2003ay}
  M.~Gockeler {\it et al.} [QCDSF Collaboration],
  Phys.\ Rev.\ D {\bf 71}, 034508 (2005).



\bibitem{Pascalutsa:2004je}
  V.~Pascalutsa and M.~Vanderhaeghen,
  Phys.\ Rev.\ Lett.\  {\bf 94}, 102003 (2005).

\bibitem{Alexandrou:2006ru}
  C.~Alexandrou, G.~Koutsou, J.~W.~Negele and A.~Tsapalis,
  Phys.\ Rev.\ D {\bf 74}, 034508 (2006).

\bibitem{Hacker:2006gu}
  C.~Hacker, N.~Wies, J.~Gegelia and S.~Scherer,
  Eur.\ Phys.\ J.\ A {\bf 28}, 5 (2006).


\bibitem{Arrington:2006zm}
  J.~Arrington, C.~D.~Roberts and J.~M.~Zanotti,
  J.\ Phys.\ G {\bf 34}, S23 (2007).










\bibitem{Pascalutsa:2007wb}
  V.~Pascalutsa and M.~Vanderhaeghen,
  Phys.\ Rev.\ D {\bf 77}, 014027 (2008).





\bibitem{Lin:2008mr}
  H.~W.~Lin and K.~Orginos,
  Phys.\ Rev.\ D {\bf 79}, 074507 (2009).



\bibitem{Alexandrou:2008bn}
  C.~Alexandrou {\it et al.},
  Phys.\ Rev.\ D {\bf 79}, 014507 (2009).

\bibitem{Alexandrou:2009hs}
  C.~Alexandrou, T.~Korzec, G.~Koutsou, C.~Lorce, J.~W.~Negele, V.~Pascalutsa, A.~Tsapalis and M.~Vanderhaeghen,
  Nucl.\ Phys.\ A {\bf 825}, 115 (2009).



\bibitem{Geng:2009ys}
  L.~S.~Geng, J.~Martin Camalich and M.~J.~Vicente Vacas,
  Phys.\ Rev.\ D {\bf 80}, 034027 (2009).






\bibitem{Cloet:2014rja}
  I.~C.~Cloet, W.~Bentz and A.~W.~Thomas,
  Phys.\ Rev.\ C {\bf 90}, 045202 (2014).


\bibitem{Shanahan:2014uka}
P.~E.~Shanahan {\it et al.} [CSSM and QCDSF/UKQCD Collaborations],
Phys.\ Rev.\ D {\bf 89}, 074511 (2014).

\bibitem{Carrillo-Serrano:2016igi}
  M.~E.~Carrillo-Serrano, W.~Bentz, I.~C.~Cloet and A.~W.~Thomas,
  Phys.\ Lett.\ B {\bf 759}, 178 (2016).


\bibitem{Bjorken:1966ij}
  J.~D.~Bjorken and J.~D.~Walecka,
  Annals Phys.\  {\bf 38}, 35 (1966).
\bibitem{Jones:1972ky}
  H.~F.~Jones and M.~D.~Scadron,
  Annals Phys.\  {\bf 81}, 1 (1973).

\bibitem{Beg:1964nm}
  M.~A.~B.~Beg, B.~W.~Lee and A.~Pais,
  Phys.\ Rev.\ Lett.\  {\bf 13}, 514 (1964).

\bibitem{Becchi:1965zz}
  C.~Becchi and G.~Morpurgo,
  Phys.\ Lett.\  {\bf 17}, 352 (1965).

\bibitem{Isgur:1981yz}
  N.~Isgur, G.~Karl and R.~Koniuk,
  Phys.\ Rev.\ D {\bf 25}, 2394 (1982).

  \bibitem{Gershtein:1981zf}
  S.~S.~Gershtein and G.~V.~Jikia,
  Sov.\ J.\ Nucl.\ Phys.\  {\bf 34}, 870 (1981)
  [Yad.\ Fiz.\  {\bf 34}, 1566 (1981)].

\bibitem{Bourdeau:1987ih}
  M.~Bourdeau and N.~C.~Mukhopadhyay,
  Phys.\ Rev.\ Lett.\  {\bf 58}, 976 (1987).

\bibitem{Gogilidze87}
S.~A.~Gogilidze, Yu.~S.~Surovtsev and F.~G.~Tkebuchava,
Sov.~J.~Nucl.~Phys. {\bf 45}, 674 (1987) [Yad. Piz. {\bf 45}, 1085
(1987)].

\bibitem{Hemmert:1994ky}
  T.~R.~Hemmert, B.~R.~Holstein and N.~C.~Mukhopadhyay,
  Phys.\ Rev.\ D {\bf 51}, 158 (1995)

\bibitem{Buchmann:2004ia}
  A.~J.~Buchmann,
  Phys.\ Rev.\ Lett.\  {\bf 93}, 212301 (2004)

\bibitem{Ramalho:2008aa}
  G.~Ramalho and M.~T.~Pena,
  J.\ Phys.\ G {\bf 36}, 115011 (2009)
\bibitem{Faessler:2006ky}
  A.~Faessler, T.~Gutsche, B.~R.~Holstein, V.~E.~Lyubovitskij, D.~Nicmorus and K.~Pumsa-ard,
  Phys.\ Rev.\ D {\bf 74}, 074010 (2006)


\bibitem{Weinberg:1978kz}
  S.~Weinberg,
  Physica A {\bf 96}, 327 (1979).


\bibitem{Butler:1993ht}
  M.~N.~Butler, M.~J.~Savage and R.~P.~Springer,
  Phys.\ Lett.\ B {\bf 304}, 353 (1993).

  \bibitem{Jenkins:1991}
E.~Jenkins and A.~V.~Manohar,
Phys.\ Lett.\ B {\bf 255}, 558 (1991).

\bibitem{Gellas:1998wx}
  G.~C.~Gellas, T.~R.~Hemmert, C.~N.~Ktorides and G.~I.~Poulis,
  Phys.\ Rev.\ D {\bf 60}, 054022 (1999).

\bibitem{Hemmert:19978}
T.~Hemmert, B.~R.~Holstein and J.~Kambor,
Phys.\ Lett.\ B {\bf 395}, 89 (1997);
  J.\ Phys.\ G {\bf 24}, 1831 (1998).

\bibitem{Arndt:2003vd}
  D.~Arndt and B.~C.~Tiburzi,
  Phys.\ Rev.\ D {\bf 69}, 014501 (2004)


\bibitem{Gail:2005gz}
  T.~A.~Gail and T.~R.~Hemmert,
  Eur.\ Phys.\ J.\ A {\bf 28}, 91 (2006)

\bibitem{Pascalutsa:2005ts}
  V.~Pascalutsa and M.~Vanderhaeghen,
Phys. Rev. Lett. {\bf 95}, 232001 (2005).

\bibitem{Kaelbermann:1983zb}
  G.~Kaelbermann and J.~M.~Eisenberg,
  Phys.\ Rev.\ D {\bf 28}, 71 (1983).

\bibitem{Bermuth:1988ms}
  K.~Bermuth, D.~Drechsel, L.~Tiator and J.~B.~Seaborn,
  Phys.\ Rev.\ D {\bf 37}, 89 (1988).

\bibitem{Lu:1996rj}
  D.~H.~Lu, A.~W.~Thomas and A.~G.~Williams,
  Phys.\ Rev.\ C {\bf 55}, 3108 (1997)


  \bibitem{Wirzba:1986sc}
  A.~Wirzba and W.~Weise,
  Phys.\ Lett.\ B {\bf 188}, 6 (1987).

\bibitem{Abada:1995db}
  A.~Abada, H.~Weigel and H.~Reinhardt,
  Phys.\ Lett.\ B {\bf 366}, 26 (1996).

\bibitem{Walliser:1996ps}
  H.~Walliser and G.~Holzwarth,
  Z.\ Phys.\ A {\bf 357}, 317 (1997).


\bibitem{Wang:2009bh}
  L.~Wang and F.~X.~Lee,
  AIP Conf.\ Proc.\  {\bf 1182}, 532 (2009).


\bibitem{Jenkins:2002rj}
  E.~Jenkins, X.~d.~Ji and A.~V.~Manohar,
  Phys.\ Rev.\ Lett.\  {\bf 89}, 242001 (2002).

\bibitem{Buchmann:2002mm}
  A.~J.~Buchmann, J.~A.~Hester and R.~F.~Lebed,
  Phys.\ Rev.\ D {\bf 66}, 056002 (2002).

  \bibitem{Leinweber:1992pv}
  D.~B.~Leinweber, T.~Draper and R.~M.~Woloshyn,
  Phys.\ Rev.\ D {\bf 48}, 2230 (1993).

  \bibitem{Alexandrou:2003ea}
  C.~Alexandrou {\it et al.},
  Phys.\ Rev.\ D {\bf 69}, 114506 (2004).

\bibitem{Alexandrou:2004xn}
  C.~Alexandrou, P.~de Forcrand, H.~Neff, J.~W.~Negele, W.~Schroers and A.~Tsapalis,
  Phys.\ Rev.\ Lett.\  {\bf 94}, 021601 (2005)

\bibitem{Ramalho:2009df}
  G.~Ramalho and M.~T.~Pena,
  Phys.\ Rev.\ D {\bf 80}, 013008 (2009)



\bibitem{Scherer:2002tk}
  S.~Scherer,
  Adv.\ Nucl.\ Phys.\  {\bf 27}, 277 (2003)
  .
\bibitem{Bernard:1995dp}
  V.~Bernard, N.~Kaiser and U.~G.~Meissner,
  Int.\ J.\ Mod.\ Phys.\ E {\bf 4}, 193 (1995).

\bibitem{Rarita:1941mf}
  W.~Rarita and J.~Schwinger,
  Phys.\ Rev.\  {\bf 60}, 61 (1941).  

\bibitem{Butler:1992pn}
  M.~N.~Butler, M.~J.~Savage and R.~P.~Springer,
  Nucl.\ Phys.\ B {\bf 399}, 69 (1993).


\bibitem{Li:2016ezv}
  H.~S.~Li, Z.~W.~Liu, X.~L.~Chen, W.~Z.~Deng and S.~L.~Zhu,
  Phys.\ Rev.\ D {\bf 95}, no. 7, 076001 (2017)
  .



\bibitem{Ecker:1994gg}
  G.~Ecker,
  Prog.\ Part.\ Nucl.\ Phys.\  {\bf 35}, 1 (1995).
\bibitem{Kim:2005gz}
  H.~C.~Kim, M.~Polyakov, M.~Praszalowicz, G.~S.~Yang and K.~Goeke,
  Phys.\ Rev.\ D {\bf 71}, 094023 (2005)
  .

\bibitem{Ramalho:2013uza}
  G.~Ramalho and K.~Tsushima,
  Phys.\ Rev.\ D {\bf 87}, no. 9, 093011 (2013)
  .

\bibitem{Dhir:2009ax}
  R.~Dhir and R.~C.~Verma,
  Eur.\ Phys.\ J.\ A {\bf 42}, 243 (2009)
  .

\bibitem{Ramalho:2013iaa}
  G.~Ramalho and K.~Tsushima,
  Phys.\ Rev.\ D {\bf 88}, 053002 (2013)
  .

\bibitem{Keller:2013hza}
  D.~Keller and K.~Hicks,
  Eur.\ Phys.\ J.\ A {\bf 49}, 53 (2013).


\bibitem{Hong:2007pr}
  S.~T.~Hong,
  Phys.\ Rev.\ D {\bf 76}, 094029 (2007)
  .

\bibitem{Jenkins:2011dr}
  E.~E.~Jenkins,
  Phys.\ Rev.\ D {\bf 85}, 065007 (2012)
  .

\bibitem{Lebed:2004fj}
  R.~F.~Lebed and D.~R.~Martin,
  Phys.\ Rev.\ D {\bf 70}, 016008 (2004)
.

\bibitem {Patrignani}
C. Patrignani et al.(Particle Data Group), Chin. Phys. C, 40, 100001
(2016).
%

\end{thebibliography}
\end{document}